\documentclass[%
 aip,
 amsmath,amssymb,
 reprint,%
]{revtex4-1}

\usepackage{graphicx}
\usepackage{dcolumn}
\usepackage{bm}

\usepackage[utf8]{inputenc}
\usepackage[T1]{fontenc}
\usepackage{mathptmx}
\usepackage{etoolbox}

\usepackage{subcaption}
\usepackage{siunitx}
\usepackage{mathastext}
\usepackage{float}

\makeatletter
\def\@email#1#2{%
 \endgroup
 \patchcmd{\titleblock@produce}
  {\frontmatter@RRAPformat}
  {\frontmatter@RRAPformat{\produce@RRAP{*#1\href{mailto:#2}{#2}}}\frontmatter@RRAPformat}
  {}{}
}%
\makeatother
\begin{document}

\preprint{AIP/123-QED}

\title{Introducing dusty plasma particle growth of nanospherical titanium dioxide}

\author{Bhavesh Ramkorun}
\author{Swapneal Jain}
\affiliation{Department of Physics, Auburn University, Auburn, AL, 36849, USA}
\author{Adib Taba}
\author{Masoud Mahjouri-Samani}
\affiliation{Department of Electrical and Computer Engineering, Auburn University, Auburn, AL, 36849, USA},
\author{Michael E. Miller}
\affiliation{Auburn University Research Instrumentation Facility, Harrison College of Pharmacy, Auburn University, Auburn, AL, 36849, USA}
\author{Saikat C. Thakur}
\author{Edward Thomas Jr.} 
\author{Ryan B. Comes}
\affiliation{Department of Physics, Auburn University, Auburn, AL, 36849, USA}

\email{ryan.comes@auburn.edu}

\date{\today}

\begin{abstract}
In dusty plasma environments, the spontaneous growth of nanoparticles from reactive gases has been extensively studied for over three decades, primarily focusing on hydrocarbons and silicate particles. Here, we introduce the growth of titanium dioxide, a wide band gap semiconductor, as dusty plasma nanoparticles. The resultant particles exhibited a spherical morphology and reached a maximum homogeneous radius of 230 $\pm$ 17 nm after an elapsed time of 70 seconds. The particle grew linearly and the growth displayed a cyclic behavior; that is, upon reaching their maximum radius, the largest particles fell out of the plasma, and a new growth cycle immediately followed. The particles were collected after being grown for different amounts of time and imaged using scanning electron microscopy. Further characterization was carried out using energy dispersive X-ray spectroscopy, X-ray diffraction and Raman spectroscopy to elucidate the chemical composition and crystalline properties of the maximally sized particles. Initially, the as-grown particles after 70 seconds exhibited an amorphous structure. However, annealing treatments at temperatures of \SI{400}{\degreeCelsius} and \SI{800}{\degreeCelsius} induced crystallization, yielding anatase and rutile phases, respectively. Notably, annealing at \SI{600}{\degreeCelsius} resulted in a mixed phase of anatase and rutile. These findings open new avenues for a rapid and controlled growth technique of titanium dioxide as dusty plasma. 

\end{abstract}

\maketitle


\begin{figure*}
    \centering
    \begin{subfigure}{0.4\linewidth}
        \includegraphics[width=\linewidth]{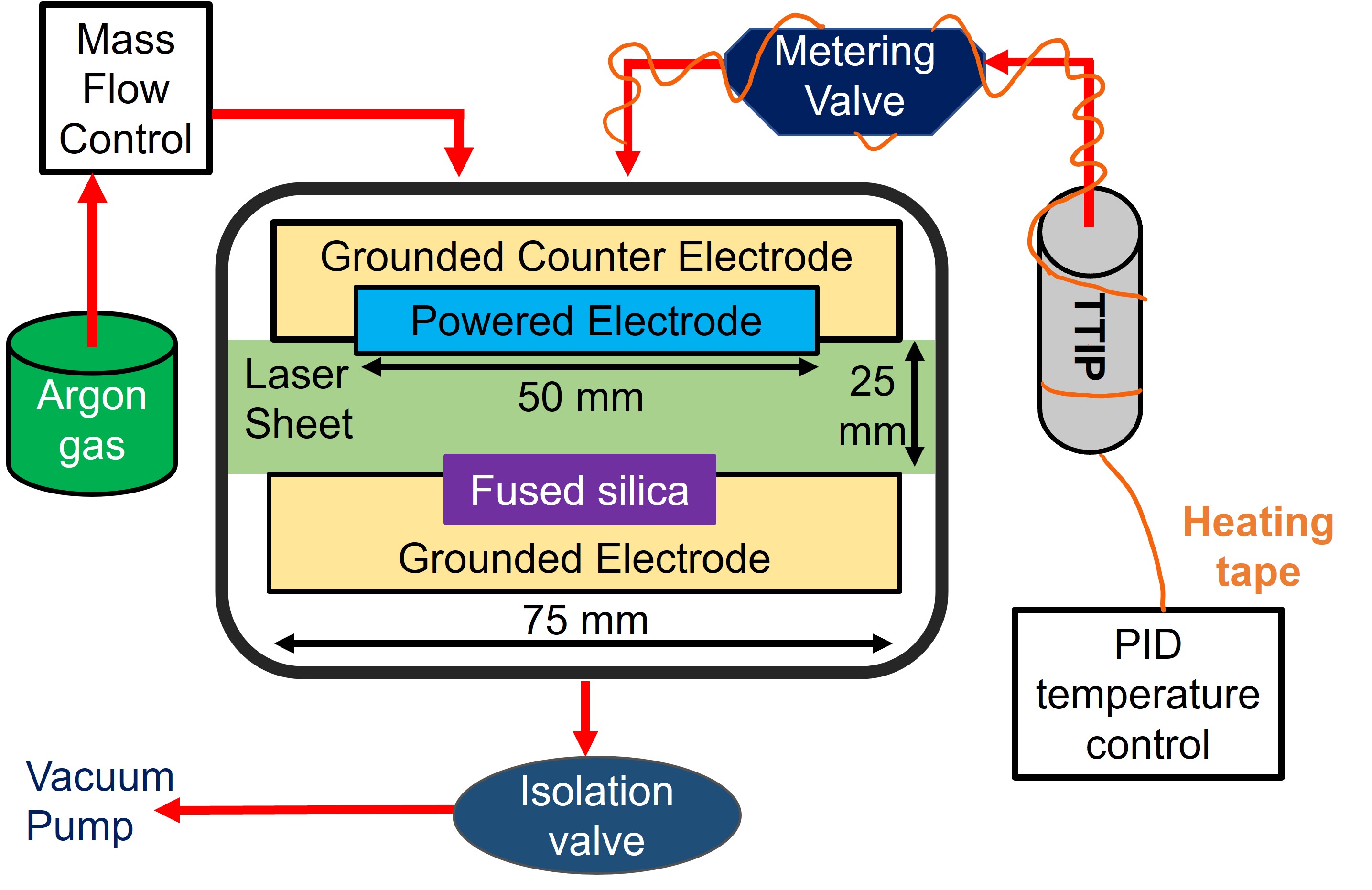}
        \caption{}
        \label{sideview}
    \end{subfigure}
    \begin{subfigure}{0.4\linewidth}
        \includegraphics[width=\linewidth]{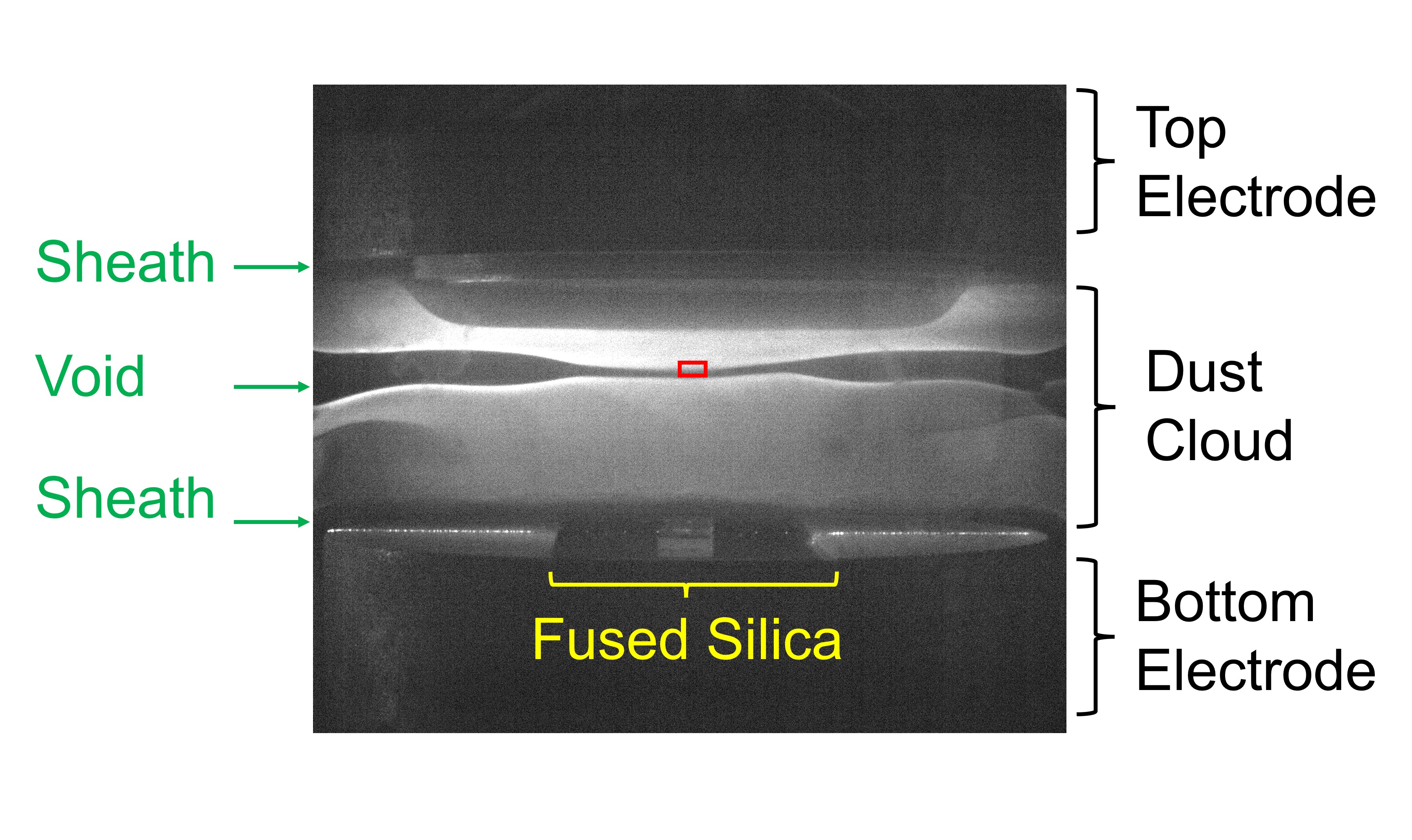}
        \caption{}
        \label{cloud_35}
    \end{subfigure}
    \begin{subfigure}{0.4\linewidth}
        \includegraphics[width=0.8\linewidth]{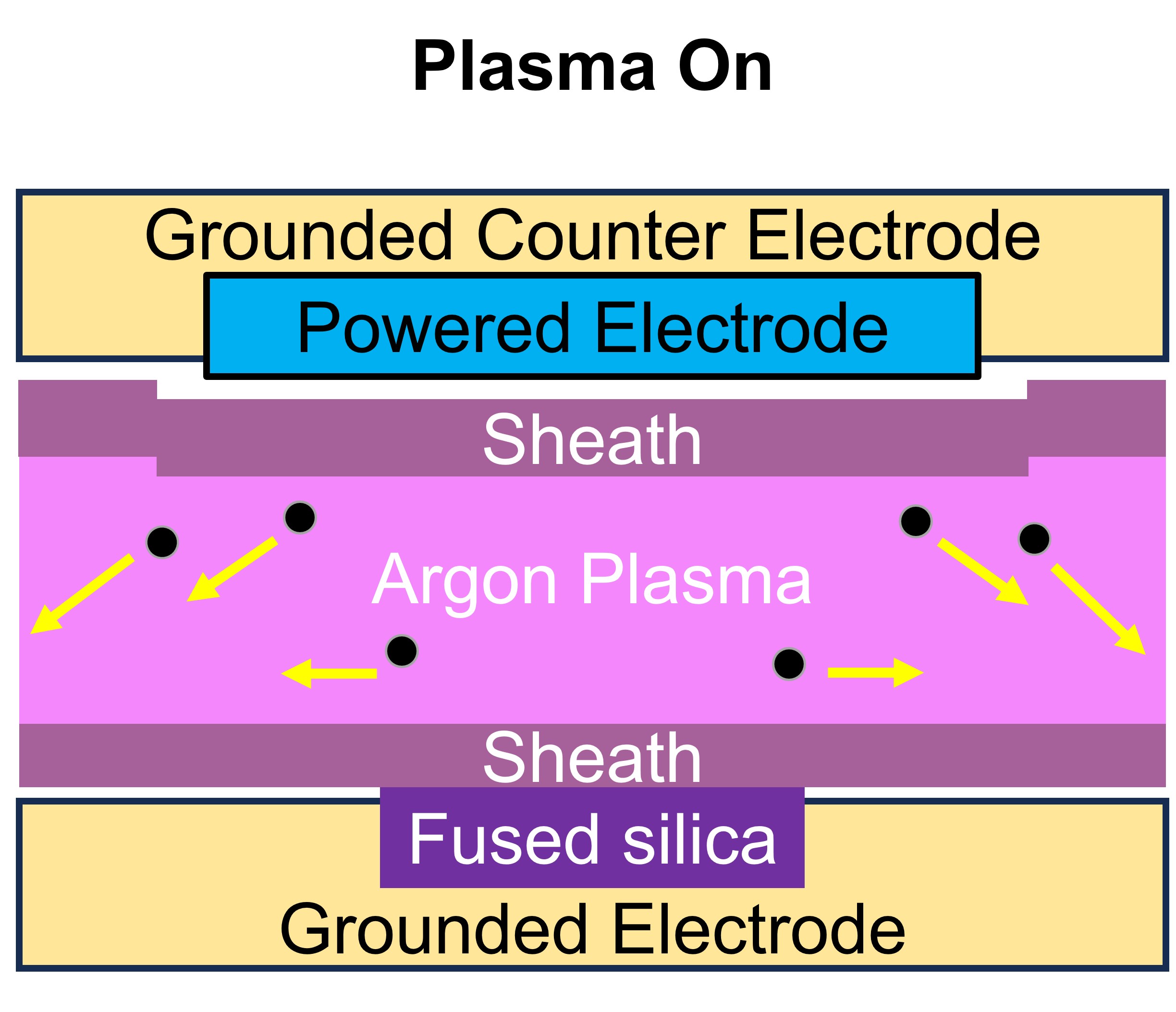}
        \caption{}
        \label{plasma_on}
    \end{subfigure}
    \begin{subfigure}{0.4\linewidth}
        \includegraphics[width=0.8\linewidth]{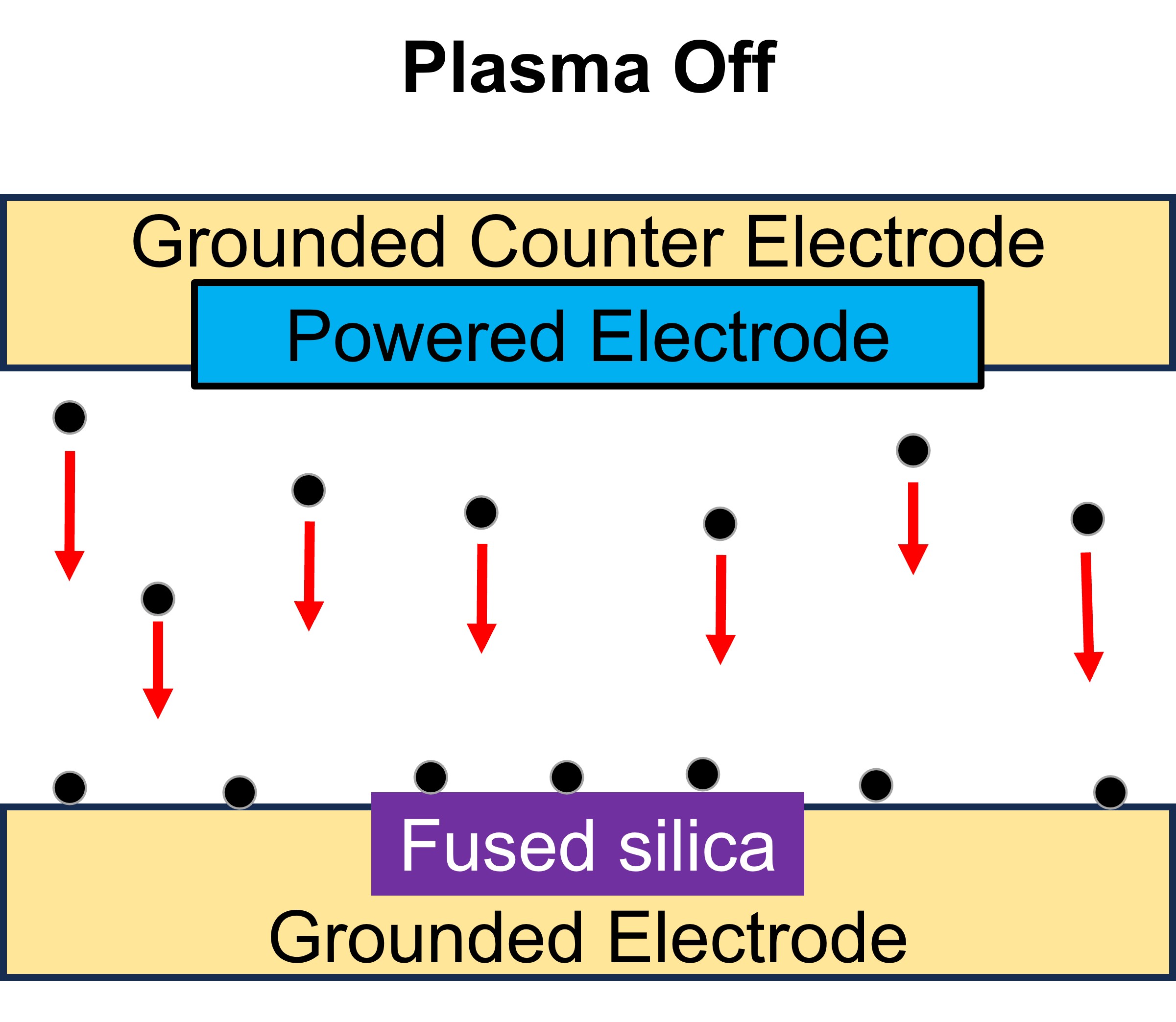}
        \caption{}
        \label{plasma_off}
    \end{subfigure}
    \caption{(a) (Not to scale) Schematic view of the experimental set up. (b) CMOS camera photograph of the dusty plasma during the first cycle showing sheath, dust, electrodes, and fused silica. (c) (Not to scale) Dust (black) deflects (yellow arrow) out of the plasma  at the end of a cycle when the plasma is on. (d) (Not to scale) Dust (black) deposits (red arrow) on the bottom electrode, including on the fused silica when the plasma is off. }
\end{figure*}

A dusty plasma refers to a plasma, the fourth state of matter, that also contains solid particles ranging in size from nanometers (nm) to micrometers. These particles are commonly referred to as "dust." These particles can spontaneously grow from reactive gases, and extensive research has delved into their formation within capacitively coupled plasmas (CCP). Much of the previous research has concentrated on the generation of either hydrocarbon or silicate dust, typically originating from acetylene $(C_2H_2)$ \cite{kovavcevic2003, kovacevic2009, kovacevic2012, couedel2019} or silane $(SiH_4)$ \cite{bouchoule1993particulate, watanabe1993growth, boufendi1994} precursors respectively.  In certain investigations, these particles have been collected and undergone detailed characterization using techniques such as deuteron-beam induced gamma-ray emission \cite{kovacevic2009}, near-edge x-ray absorption fine structure spectroscopy (NEXAFS) \cite{kovacevic2012}, and Raman spectroscopy \cite{couedel2019}. Furthermore, observations utilizing  scanning electron microscopy (SEM) \cite{kovavcevic2003}  and transmission electron microscopy (TEM) \cite{boufendi1994} have consistently revealed that these particles exhibit spherical morphology and their radii exhibited linear growth characteristics over time. Moreover, studies have shown that dusty plasma particle growth processes offer precise control over nanoparticles' size and morphology  \cite{ganguly1993growth, garscadden1994overview, cao2002deposition, galli2009charging, chutia2021nanodusty}. 

These investigations have spurred the development of more scientifically valuable particulates in dusty plasma systems, such as semiconductors and quantum dots \cite{kortshagen2009nonthermal, boufendi2011}. Over the last decade, organosilicon, conductive polymers and metallic particles have been cultivated within CCP dusty plasma environments using hexamethyldisiloxane \cite{despax2012}, aniline \cite{pattyn2018} and aluminium chloride \cite{cameron2023capacitively} respectively. These precursors have traditionally found application in plasma-enhanced chemical vapor deposition (PECVD) for thin film production \cite{wang1997dependence, Shirafuji99, Shioya05, airoudj2008}. 

This letter introduces semiconducting titanium dioxide ($TiO_2$) particle growth within a dusty plasma environment. The growth process was initiated using titanium (IV) isopropoxide (TTIP) $\left(Ti\left(OC_3H_7\right)_4 \right)$ (Sigma Aldrich 687502-25G)  as the reactive precursor. X-ray diffraction (XRD) and Raman spectroscopy were employed to measure the crystal phases. SEM was used to image and subsequently measure the sizes of the particles grown as a function of time. SEM energy-dispersive X-ray spectroscopy (EDS) is used to ascertain the chemical composition of the particles. This investigation not only presents a novel growth technique for $TiO_2$ particles, but also introduces the material for further study in the field of dusty plasmas.

The use of TTIP has had various applications for the production of $TiO_2$. A consistent observation in these investigations is the necessity of an annealing process to promote crystallization. In the case of PECVD, thin films exhibited a relatively slow growth rate, typically ranging from 20 to 80 nm per hour \cite{Lee94, Yang06}. Furthermore, these films required annealing at temperatures exceeding \SI{400}{\degreeCelsius} to crystallize. Thin films were also grown using atomic layer deposition. Whereas one study achieved crystallization on substrates heated to \SI{250}{\degreeCelsius} \cite{aarik2000titanium}, another study annealed at \SI{900}{\degreeCelsius} to attain crystallization \cite{lee2013deposition}. Chemical spray pyrolysis was also used to produce $TiO_2$, and the samples needed annealing beyond \SI{700}{\degreeCelsius} to crystallize \cite{juma2015effect}. Sol-gel methods for synthesizing $TiO_2$, a process that lasted two days, also required annealing at \SI{700}{\degreeCelsius} to induce crystallization \cite{muaz2016effect}.

TTIP has also been employed in molecular beam epitaxy (MBE), in combination with an oxygen plasma, to achieve a growth rate of 125 nm per hour for $TiO_2$ thin films on substrates heated to temperatures up to \SI{725}{\degreeCelsius} \cite{jalanTiO2}. Similarly, TTIP has been instrumental in growing $SrTiO_3$ in MBE on substrates heated to temperatures ranging from 900 to \SI{1100}{\degreeCelsius}, with documented growth rates exceeding 100 nm per hour in some instances \cite{jalanSTO, thapaSTO, engel-herbertSTO}. Dusty plasma particle growth, however, is not limited by substrate constraints; the dust particles formed in the plasma can be collected on any desired substrate at room temperature.

Nucleation of nanoparticles and the formation of spherical aggregates are essential prerequisites for material growth. This phenomenon has been well-documented not only for TiO$_{2}$ thin films \cite{huang2002comparison, Nguyen13}, but also for various films containing carbon \cite{Stoner92, Kuhr95, Ramkorun21}. Dusty plasma has already exhibited rapid nucleation of nm-sized carbon particles within milliseconds \cite{ravi2009}. These particles can subsequently grow to sizes of hundreds of nm within a matter of seconds \cite{groth2015}. Here we demonstrate that dusty plasma processes can grow $TiO_2$ nanospheres within 10 seconds. These spheres continue to grow linearly until 77 seconds. Although the as-grown particles collected after 70 seconds were initially amorphous, they crystallized into either anatase or rutile upon annealing. According to literature, rutile exhibits a direct wide band gap and anatase exhibits an indirect wide band gap  \cite{zhang2014new}. They both have scientific applications as photocatalysts \cite{luttrell2014anatase}.


Figure \ref{sideview} provides a schematic representation of the experimental configuration. The plasma chamber, a 6-way stainless ISO-100 cross, contains a pair of electrodes, each with a 75 mm diameter and separated by a distance of 25 mm. On the top electrode, a 50 mm segment was designated as the powered region, while the remainder functioned as a grounded counter electrode, with separation facilitated by an insulating teflon ring. These electrodes had thicknesses of 13 mm (top) and 19 mm (bottom). Notably, a 1 mm deep slot was incorporated into the surface of the bottom electrode, positioned at its midpoint, where a fused silica slide measuring $75 mm \times 25 mm$ was precisely fitted. This slide serves as a collection surface for the nanoparticles.

The base chamber pressure was $3 \pm 0.3$ millitorr (mTorr). The line connecting TTIP to the chamber was wrapped with heating tape (Omega Engineering, Inc SRT052-120) and heated to a steady temperature of \SI{75}{\degreeCelsius} using a PID temperature controller (Inkbird 40A Solid State Relay). This setup allowed for a regulated flow of TTIP vapor into the chamber, which was precisely controlled by means of a low-flow metering valve (Swagelok SS-SVR4-VH). When the valve was opened, the chamber pressure increased to 35 $\pm$ 3 mTorr. Subsequently, argon (Ar) gas was introduced into the chamber via a mass flow controller, set to deliver a flow rate of 7 standard cubic centimeters per minute. This raised the total chamber pressure to 45 $\pm$ 3 mTorr. Finally, to attain a desired total chamber pressure of 300 $\pm$ 1 mTorr, an isolation valve, connecting the chamber to the vacuum pump, was incrementally closed.

The plasma was ignited using a fixed frequency RF generator operating at 13.56 MHz with a power output of 300 W (RF VII, Inc RF-3-XIII). An auto-matching network (RF VII, Inc AIM-5/10) was employed to optimize the system. The forward power was set to 30 W, while the reflected power remained at a minimal 1 W, enabling the controlled growth of a dust cloud within the plasma environment. The dust particles were suspended between the electrodes by a balance of forces, including gravitational force, electric force due to the capacitively coupled electrodes, thermophoretic force, ion drag force, and neutral drag force, as described in previous studies \cite{merlino06, van2015}. As the dust particles grew larger, the gravitational force gradually became the dominant factor, causing the particles to no longer remain suspended in the plasma. The particles in this experiment reached this radius after 77 seconds, after which a new growth cycle immediately followed.

\begin{figure}[ht]
    \centering
    \includegraphics[width=0.5\textwidth]{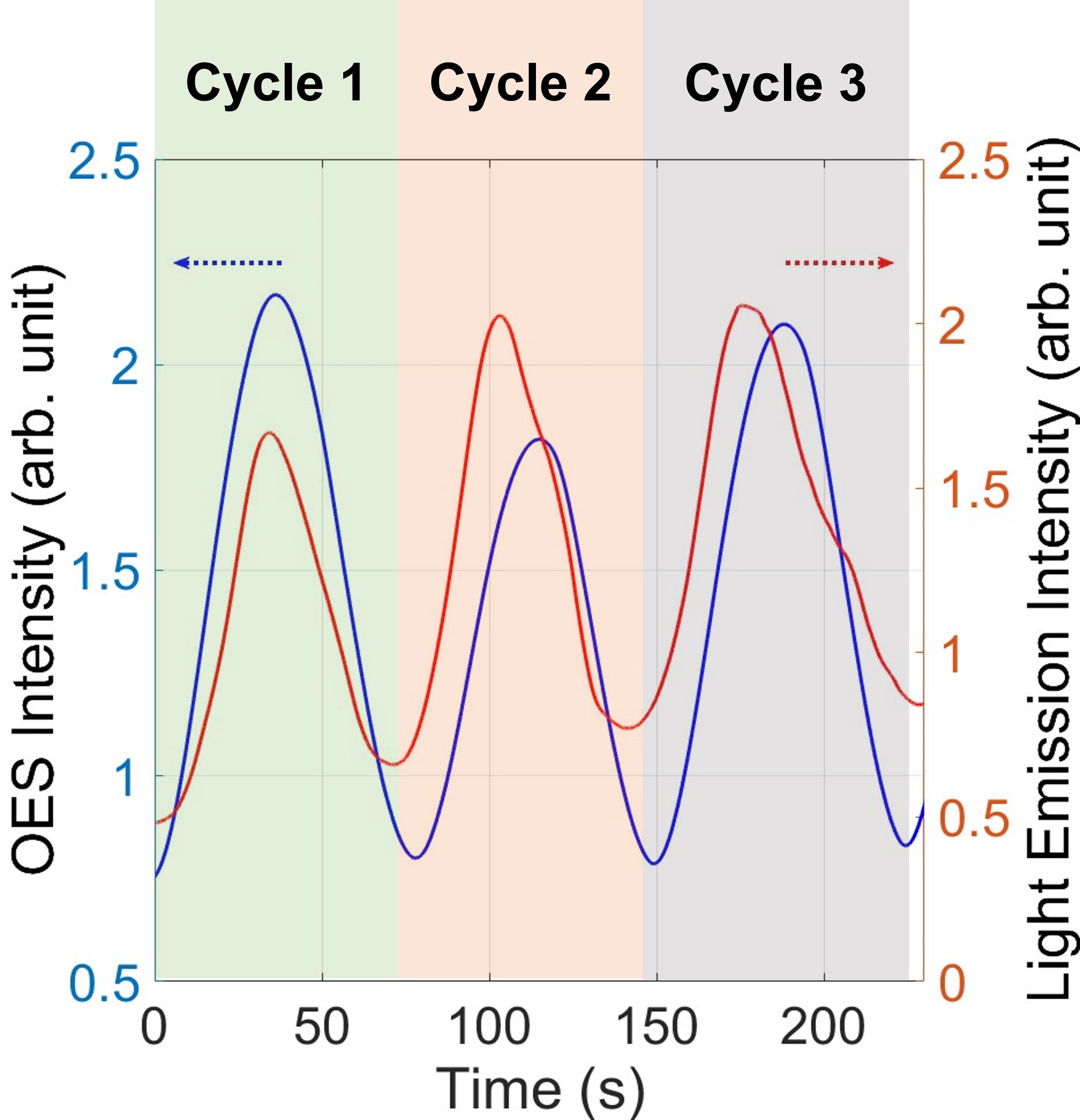}
    \caption{ OES and light emission intensity showing the dusty plasma's cyclic behavior. Light emission intensity was calculated from red region of Fig \ref{cloud_35} over all the frames recorded for three cycles.}
    \label{cycle}
\end{figure}

To visualize the dust cloud in two dimensions, a green laser sheet with a wavelength of 532 nm was directed into the plasma. Imaging of the dust cloud, perpendicular to the laser sheet, was achieved using a complementary metal oxide semiconductor (CMOS) camera (xiQ MQ042MG-CM) at 50 frames per second (FPS). In Figure \ref{cloud_35}, an illustrative depiction of a dust cloud is presented. Notably, there was a recurrent appearance of a void, a dust-free region, in the central part of the plasma. This void exhibited cyclic expansion and contraction, mirroring the particle growth cycle. To quantify this cycle, the light emission intensity of a region, highlighted in red was measured. Concurrently, optical emission spectroscopy (OES) was employed to monitor the plasma throughout the particle growth process. A broadband spectrometer (Avantes ULS4096CL) was utilized, boasting a resolution of 0.59 nm, a 25 micrometer slit size, a 600 lines/mm grating. Five data points, each with 100 millisecond integration time were averaged. The cyclic evolution of intensity of Ar I line at 763.5 nm, corresponding to the electric dipole transition of meta-stable Ar from $3s^23p^5\left(^2P^0_{3/2}\right)4p$ to $3s^23p^5\left(^2P^0_{3/2}\right)4s$, was chosen as the indicator of the growth cycle \cite{norlen1973, wiese1989, NIST_ASD}. The resulting data, presented in Figure \ref{cycle}, revealed that the growth cycle duration was approximately 77 $\pm$ 9 s and 77 $\pm$ 4 s, according to light emission intensity and OES respectively.

In addition to the light emission and OES measurements, a size analysis was performed on the grown particles using SEM.  To perform these measurements, it was necessary to develop a technique that reliably captured the grown particles at specific times. The plasma was activated for various duration, and upon deactivation, dust particles deposited onto the fused silica, where they were gathered for subsequent analysis.
Generally, when the dusty particles reached their maximum size, the primary forces influencing them are electric and gravitational forces \cite{van2015}. When the plasma is still active, a sheath layer (i.e., a region in the plasma near a surface which the separation between the ion and electron densities leads to formation of an electric field) forms above the silica slide. The negatively charged dust particles, once reaching a critical charge-to-mass ratio, will interact with the sheath electric field and be deflected away from the region between the electrodes towards the edges of the plasma volume, as depicted in Fig. \ref{plasma_on}. However, if the plasma is extinguished, the sheath electric field is no longer present and the formed particles will fall directly towards the bottom electrode and ultimately the surface of the glass slide under the influence of gravity, as depicted in Fig. \ref{plasma_off}. This enables unambiguous collection and measurement of the particle size as a function of time.

\begin{figure}[ht]
    \centering
    \begin{subfigure}{0.5\textwidth}
        \includegraphics[width=\textwidth]{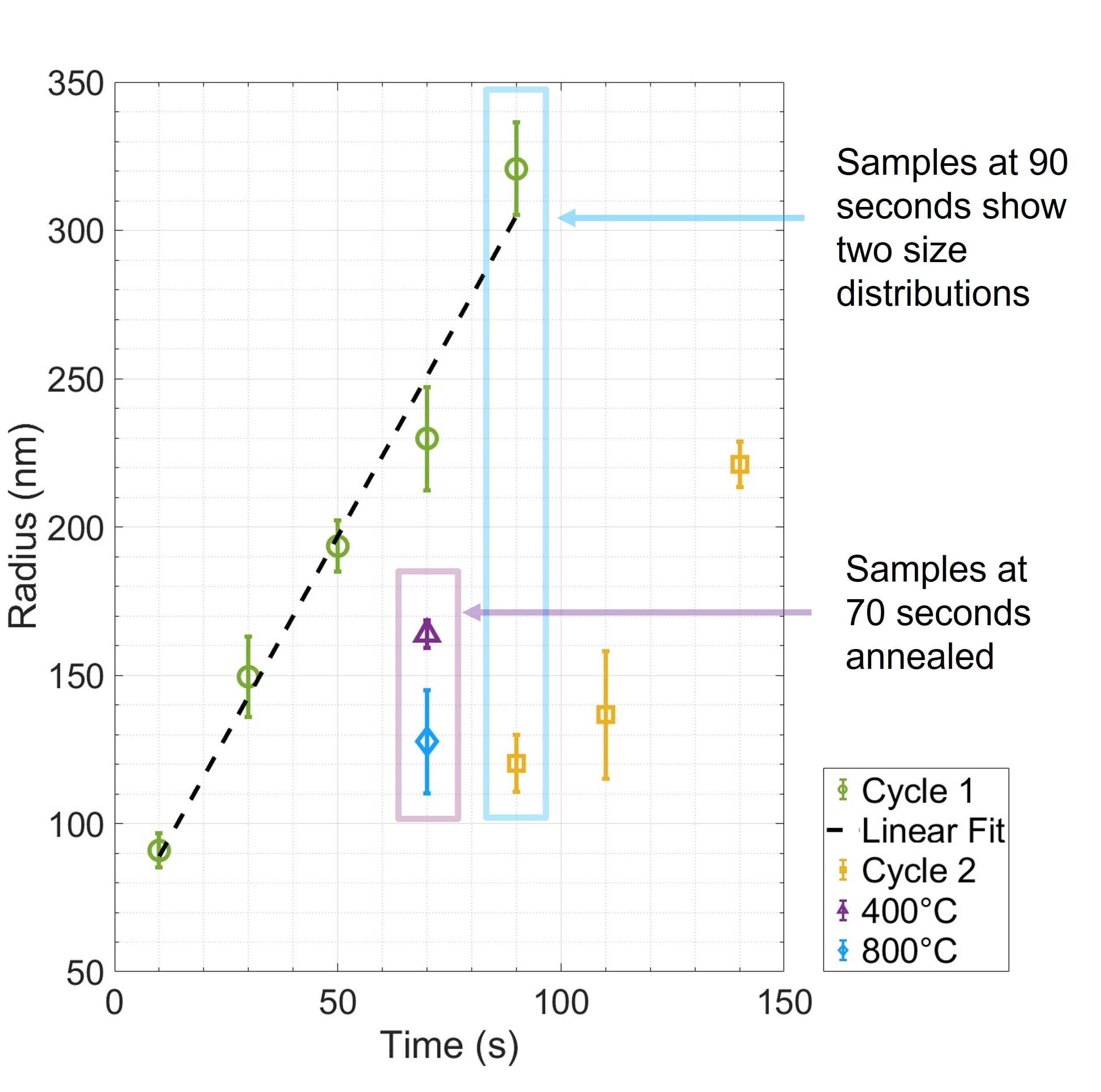}
        \caption{}
        \label{linear_grow}
    \end{subfigure}
    \begin{subfigure}{0.5\textwidth}
        \includegraphics[width=0.8\textwidth]{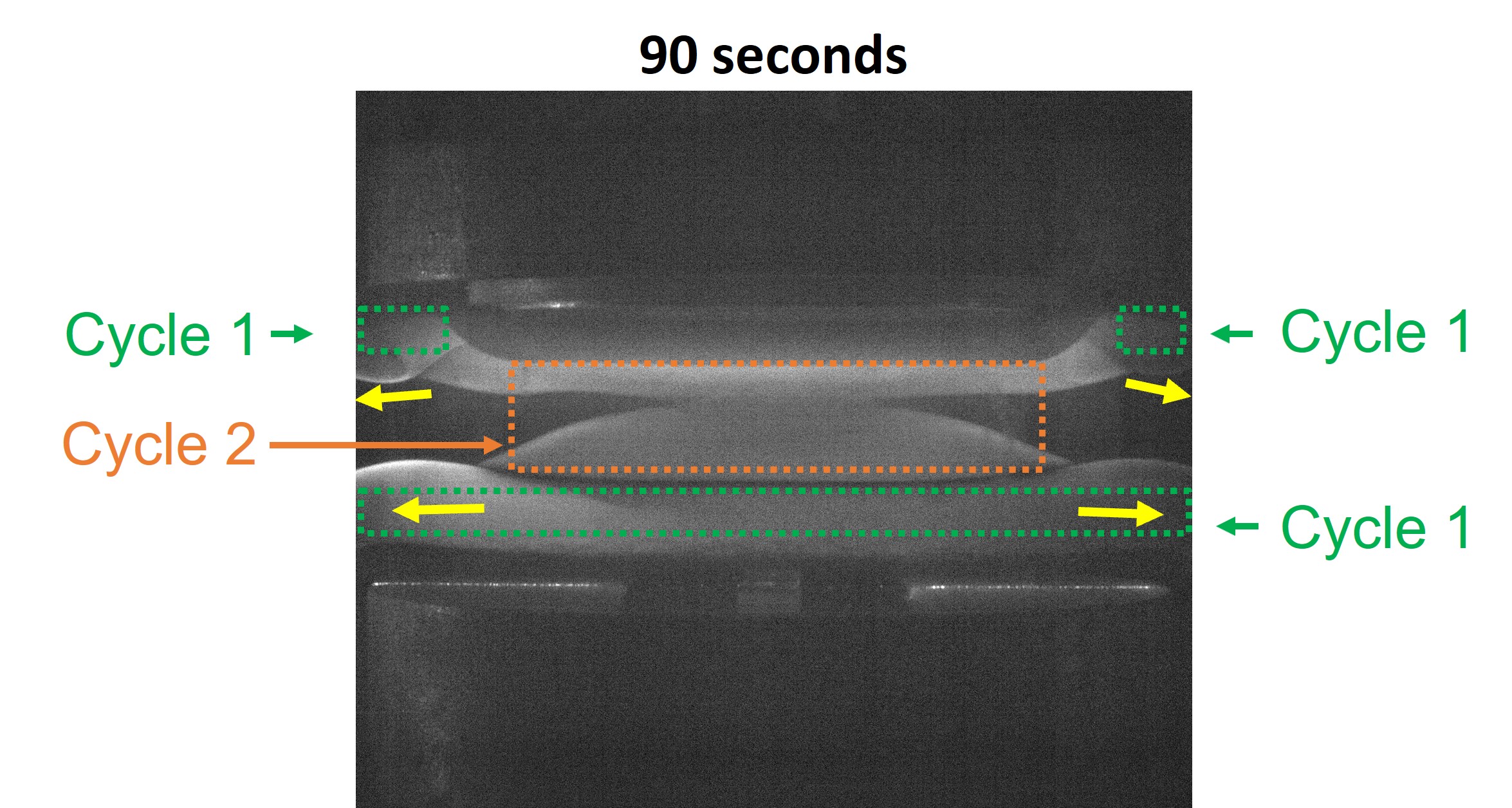}
        \caption{}
        \label{cloud90}
    \end{subfigure}
    \caption{(a) The particle size distribution over time. A linear fit (dashed line) is applied to the data points between 10 and 90 seconds. At 70 seconds, three data points are shown. A radius of 229 $\pm$ 17 nm corresponds to the as-grown particle within the first cycle. Radii of 163 $\pm$ 4 nm and 127 $\pm$ 17 nm correspond to samples annealed at 400 and \SI{800}{\degreeCelsius}, respectively.  At 90 seconds, two data points are observed, reflecting dust collected from two cycles. Radii of 320 $\pm$ 15 nm and 120 $\pm$ 9 nm correspond to the cycle 1 and 2, respectively. 
    (b) Dust cloud at 90 seconds shows two cycles. Cycle 1 and 2 boxed in green and orange, respectively. Yellow arrows indicate dust from cycle 1 deflecting away, similar to Fig. \ref{plasma_on}. After plasma deactivation, two different dust sizes were collected.}    
\end{figure}

Prior to SEM, samples were sputter coated with gold using an EMS Q150R Sputter Coating Device (Electron Microscopy Sciences) and imaged at 20 kV using a Zeiss EVO50 micoscope (Carl Zeiss Microscopy, LLC). In Figure \ref{linear_grow}, a linear progression in particle size is observed as a function of plasma runtime, until the culmination of each growth cycle. Subsequently, after a duration of approximately 77 seconds, a new cycle commenced, marking the onset of a new phase of particle generation and growth. At 90 seconds, some dust from the first cycle are still flowing out of the plasma, as shown in Fig. \ref{cloud90}. Therefore, there are two size distribution. However, there is no dust from the first cycle on the samples collected beyond 90 seconds. Linear growth during the first cycle was evident for samples collected after 10 seconds. The first 10 seconds is dominated by creation of radicals, ions and species in the nucleation stage from the initial gas molecules. Because of the non-linear dynamics in this stage, we only consider the subsequent growth after 10 seconds.

\begin{figure}[ht]
  \centering
  \begin{subfigure}{0.4\textwidth}
    \includegraphics[width=\linewidth]{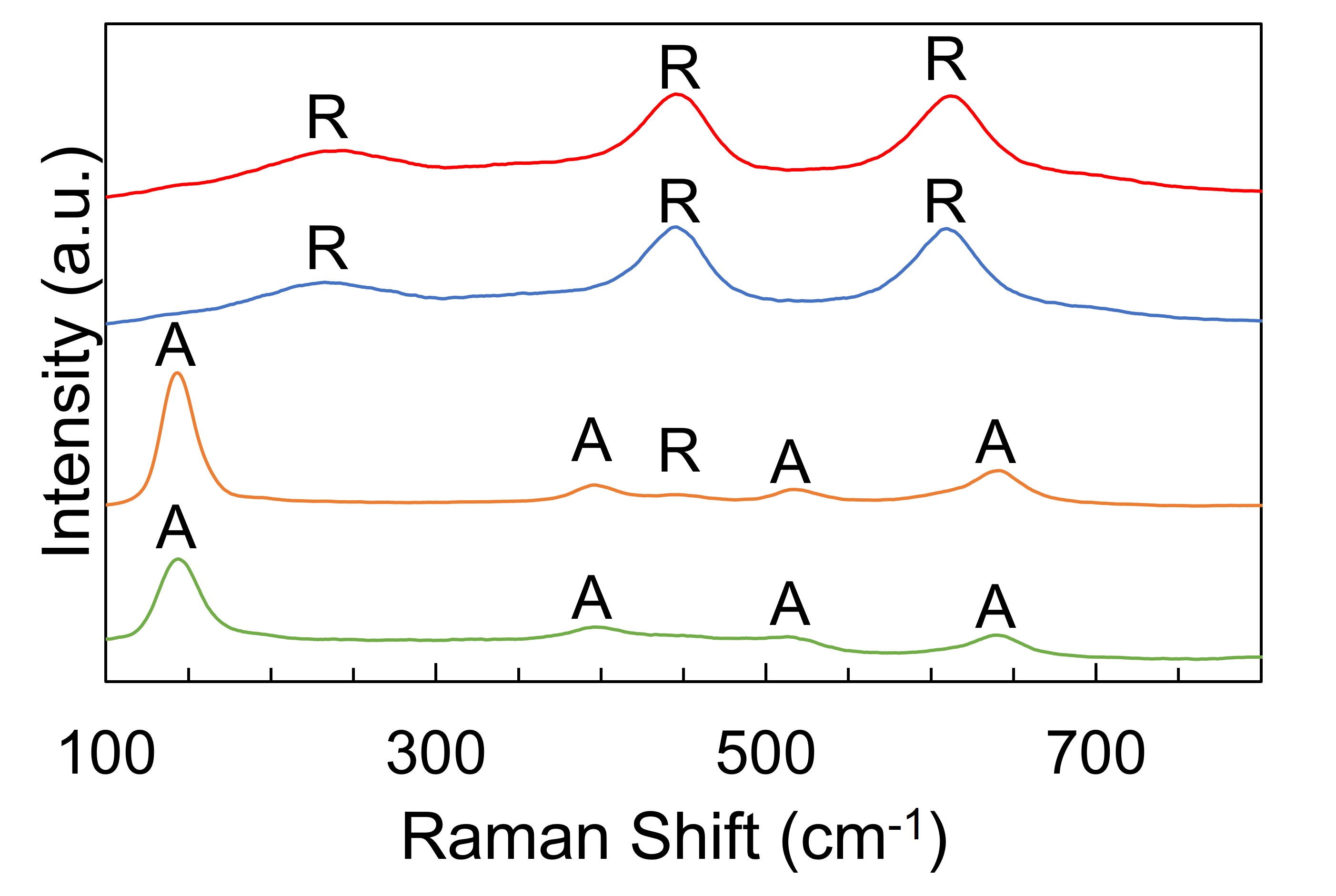}
    \caption{}
    \label{Raman}
  \end{subfigure}
  \hfill
  \begin{subfigure}{0.4\textwidth}
    \includegraphics[width=\linewidth]{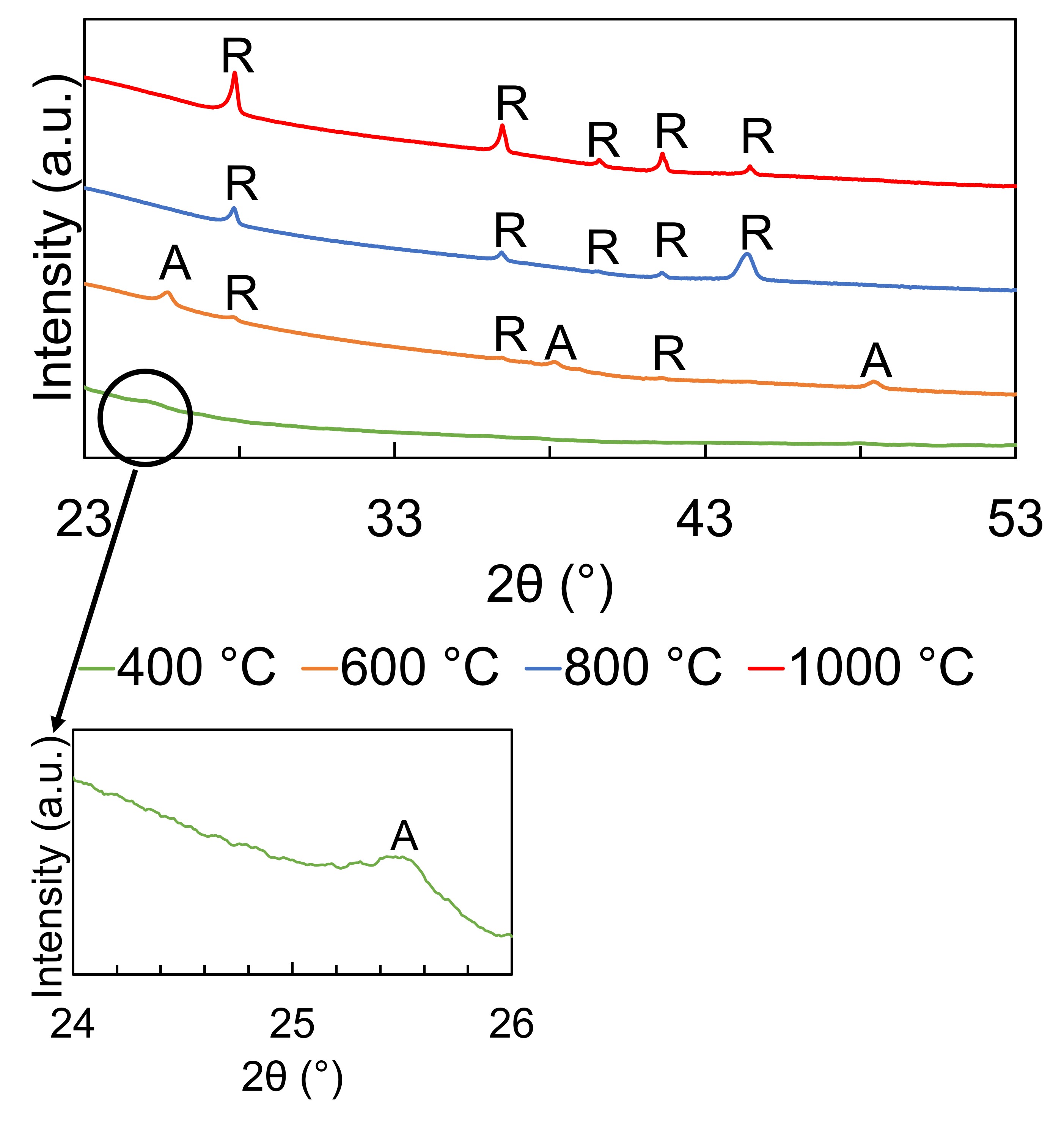}
    \caption{}
    \label{XRD}
  \end{subfigure}

  \caption{[A: Anatase; R: Rutile] Characterization of the particles grown for 70 seconds and annealed for 2 hours. (a) Raman spectroscopy and (b) XRD respectively, showing anatase at \SI{400}{\degreeCelsius} and rutile at \SI{800}{\degreeCelsius} and \SI{1000}{\degreeCelsius}.  At \SI{600}{\degreeCelsius} there is a mixed phase of anatase and rutile. There is an anatase peak on the XRD of \SI{400}{\degreeCelsius} as shown magnified.}
  \label{Characterisation}
\end{figure}

The 70 second particles were collected twenty times on the same substrate to increase particle density for the purpose of material analysis. The plasma was pulsed twenty times through activation for 70s and deactivation for 45 seconds. For individual samples, a two-hour annealing process was conducted at four different temperatures: 400, 600, 800, and \SI{1000}{\degreeCelsius}. At \SI{400}{\degreeCelsius}, $TiO_2$ underwent crystallization, resulting in the formation of anatase, while at \SI{800}{\degreeCelsius}, the transformation into rutile occurred. At \SI{600}{\degreeCelsius}, a mixed phase of anatase and rutile was observed.  The Raman spectra, as displayed in Fig. \ref{Raman}, closely resembled the known patterns documented in literature for anatase and rutile \cite{balachandran1982}. Additionally, XRD, as illustrated in Fig. \ref{XRD}, confirmed that the particles matched the known powder diffraction pattern for each phase. As mentioned above, annealing is generally required to cryatallize $TiO_2$ grown from TTIP. We also found it necessary to anneal our samples to achieve crystallization. However, when contrasted with the different aforementioned growth techniques, dusty plasma offers clear benefits, such as the ability to achieve nanoparticle formation within just 10 seconds, without being constrained by substrate limitations, and providing precise control over spherical dimensions through a linear growth rate.

\begin{figure*}[ht]
    \centering
    \begin{subfigure}[b]{0.45\textwidth}
        \includegraphics[width=\textwidth]{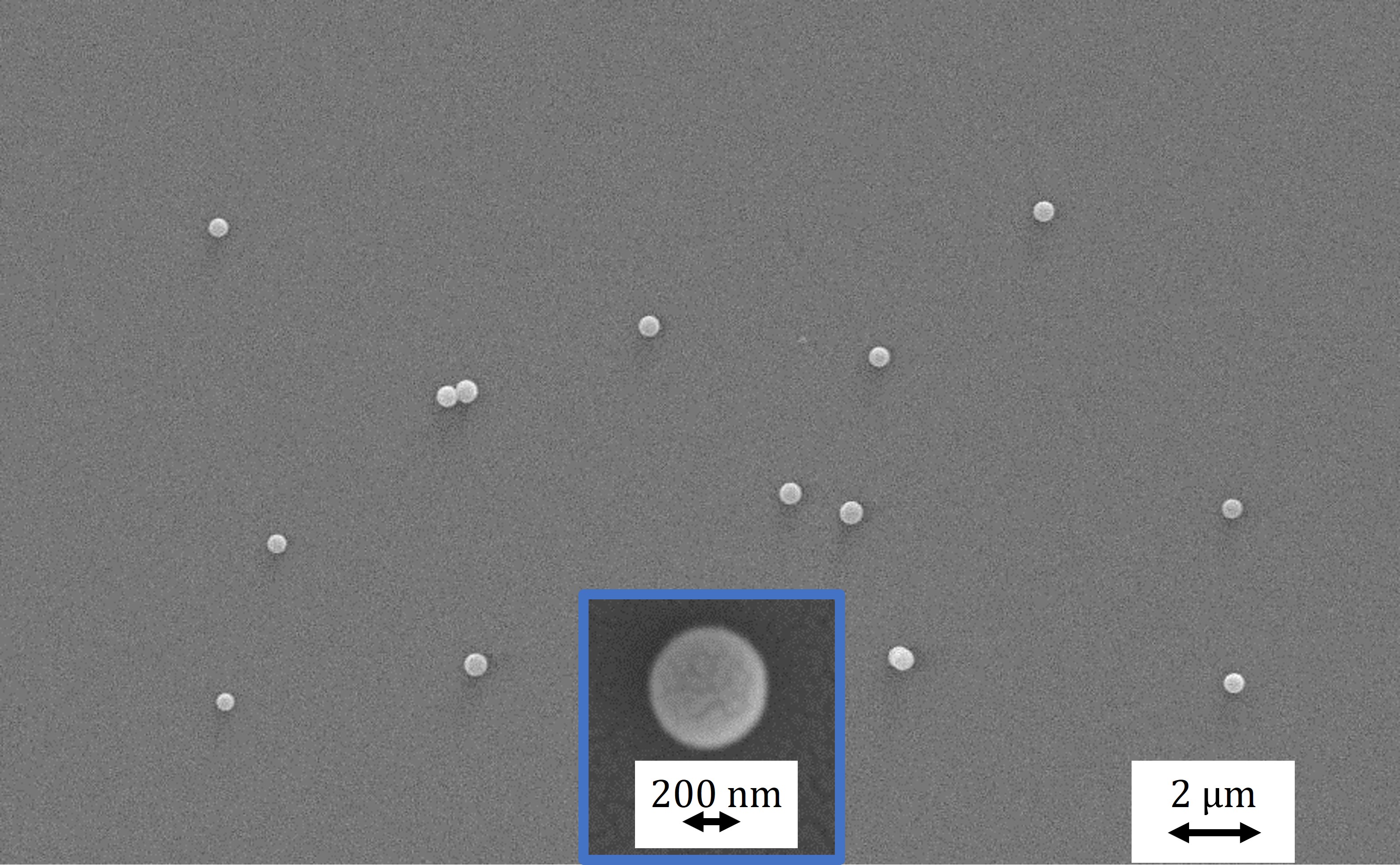}
        \caption{}
        \label{subfig:a}
    \end{subfigure}
    \hfill
    \begin{subfigure}[b]{0.45\textwidth}
        \includegraphics[width=\textwidth]{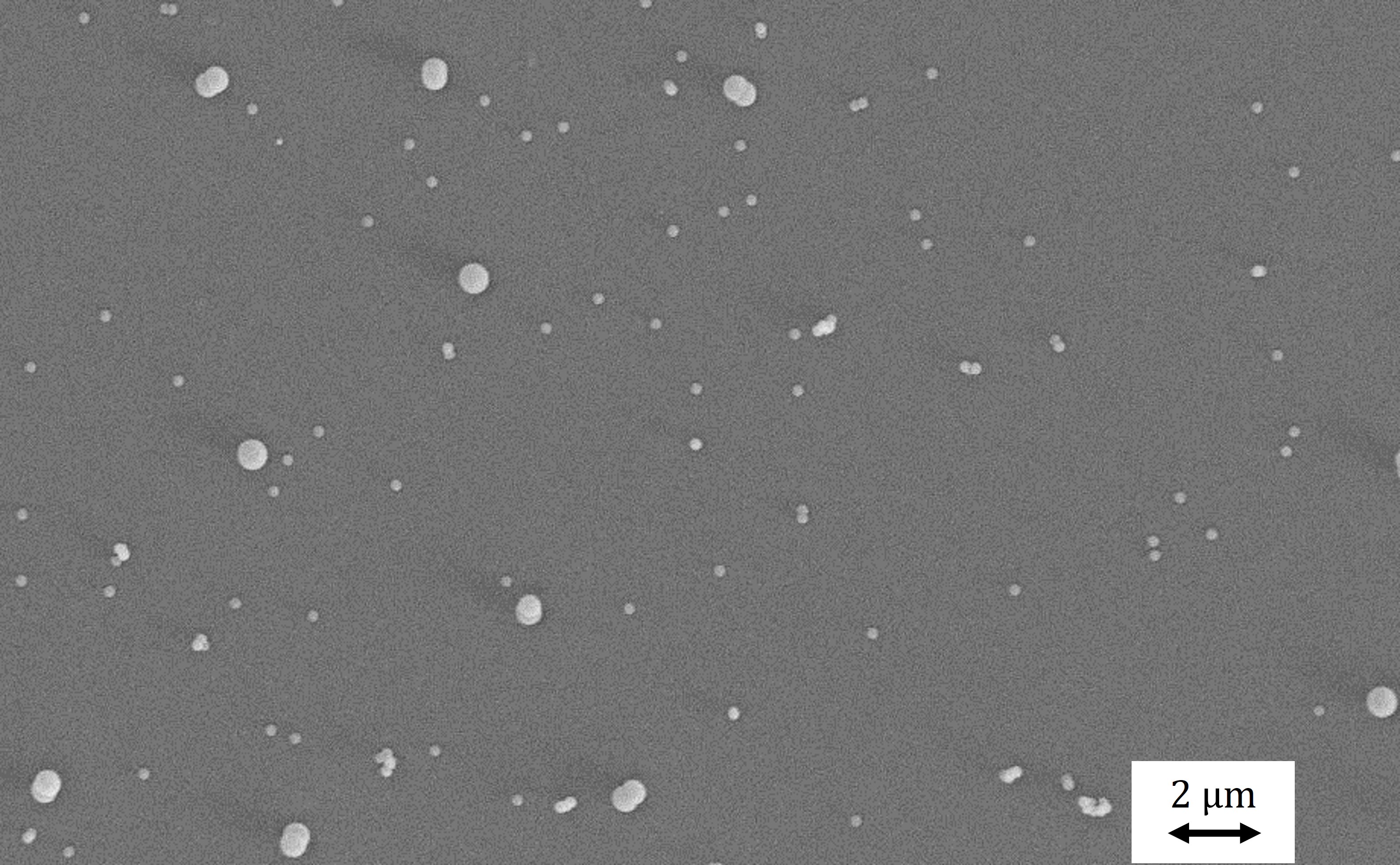}
        \caption{}
        \label{subfig:b}
    \end{subfigure}    
    \vspace{\baselineskip} 
    \begin{subfigure}[b]{0.45\textwidth}
        \includegraphics[width=\textwidth]{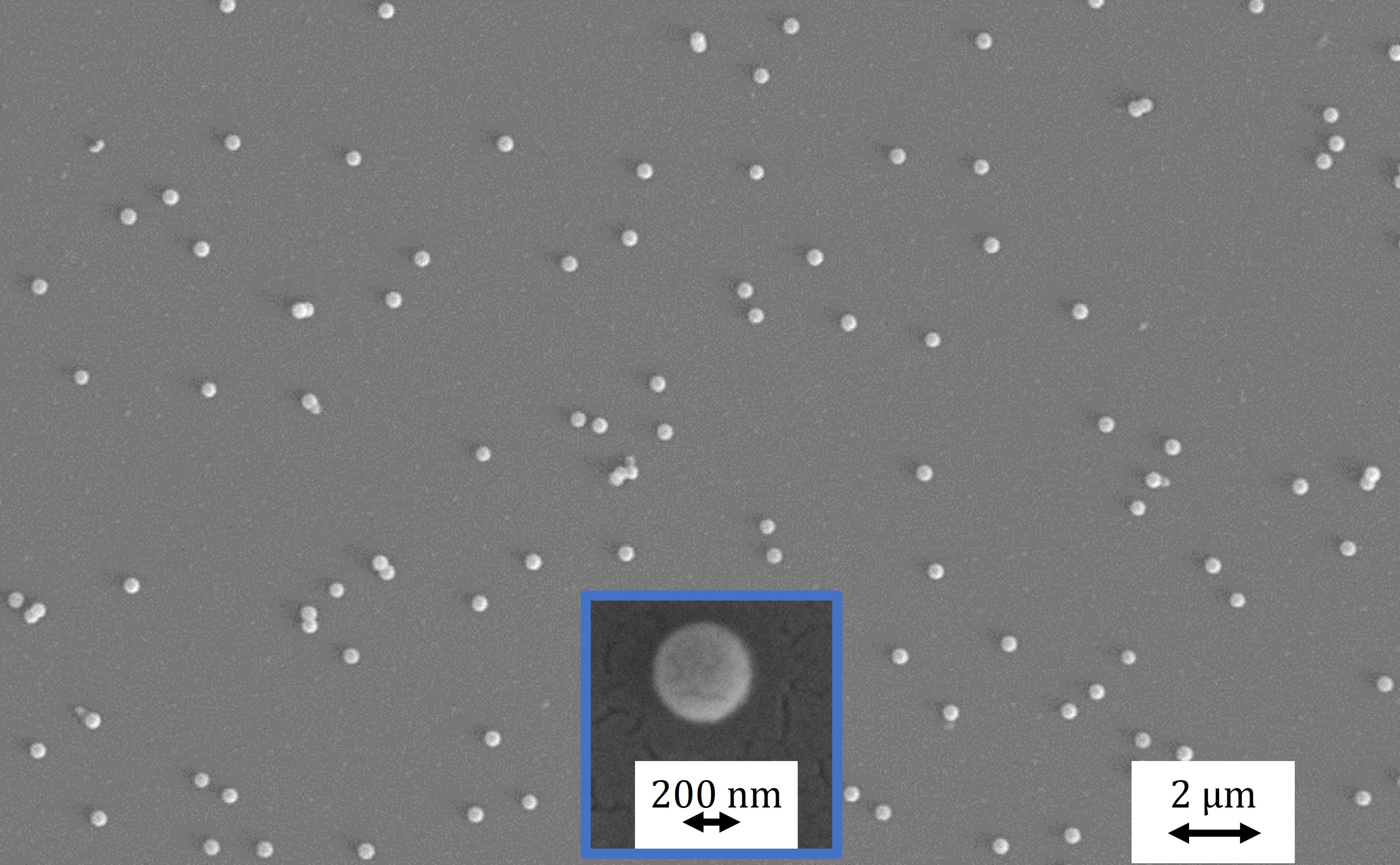}
        \caption{}
        \label{subfig:c}
    \end{subfigure}
    \hfill
    \begin{subfigure}[b]{0.45\textwidth}
        \includegraphics[width=\textwidth]{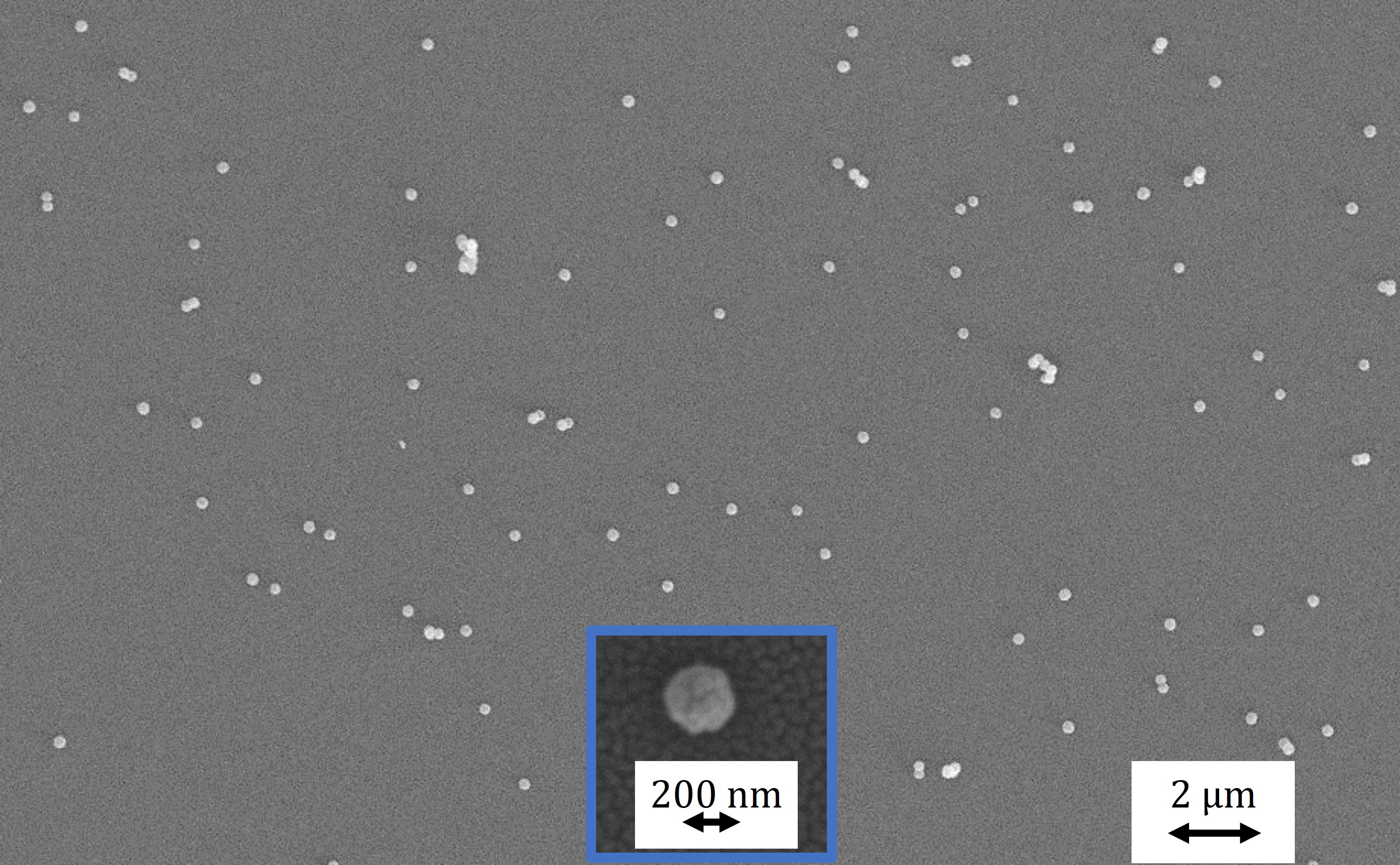}
        \caption{}
        \label{subfig:d}
    \end{subfigure}    
    \caption{SEM images of nanoparticles collected on fused silica after deactivating plasma. There is a gradual decrease in radii with higher annealing temperatures. (a) As-grown after 70 seconds. (b) As-grown after 90 seconds, showing two different sizes  (c) After 70 seconds and annealed to \SI{400}{\degreeCelsius} (Anatase). (d) After 70 seconds and annealed to \SI{800}{\degreeCelsius} (Rutile).  }
    \label{SEM}
\end{figure*}

SEM images, illustrated in Figure \ref{SEM}, were also employed to examine the particles' size before and after annealing. These images clearly indicated a reduction in mean particle radius following the annealing process. Particles that had been grown for 70 seconds prior to annealing exhibited an average radius of 230 $\pm$ 17 nm, while particles subjected to a 2-hour annealing treatment at 400 and \SI{800}{\degreeCelsius} displayed a reduced average radius of 164 $\pm$2 nm and 128 $\pm$ 17 nm, respectively. SEM of a 90 second sample, with two size distributions is also shown. Additionally, quantitative EDS analysis was performed using an Oxford INCA EDS analysis system (Oxford Instruments America, Inc.). The findings, as summarized in Table \ref{EDS}, showed a decline in the weight percentage of carbon (C) and an increase in titanium (Ti) and oxygen (O) on the samples as the annealing temperature increased. The combined SEM and EDS results suggest that C was oxidized and desorbed from the nanoparticle during annealing, resulting in reduced C concentration and smaller particle radii.  

\begin{table}
\caption{Summarizing the impact of annealing on weight (\%) of various elements grown for 70 seconds, as determined by EDS. Presence of Si is due to the fused silica substrate. There is also a decrease in radii of the nanoparticles between the as-grown and the annealed as shown in Fig. \ref{linear_grow}. Fig. \ref{SEM} shows SEM images of the samples.}
\centering
\begin{tabular}{cccc}
\hline
\hline
Element & \multicolumn{3}{c}{Weight (\%)} \\
\cline{2-4}
& As-grown & \SI{400}{\degreeCelsius} & \SI{800}{\degreeCelsius} \\
\hline
C K & 32.32 & 2.45 &  1.62 \\
O K & 27.39 & 47.83 & 49.35 \\
Si K &  32.87 & 41.11 & 36.45 \\
Ti K & 7.42 & 8.60 & 12.57 \\
\hline
\hline

\end{tabular}
\label{EDS}
\end{table}

TTIP decomposed into molecules within the plasma, providing the source of Ti, O, and C in the nanoparticles. The fused silica substrates are the source of the Si signal and additional O. In literature, non-invasive diagnostic methods were employed to ascertain the breakdown of TTIP. For instance, in ultrahigh vacuum ranging from $1\times 10^{-9} - 1\times 10^{-5}$ torr during PECVD, molecular beam scattering and temperature programmed reaction spectroscopy identified the conversion of TTIP into $TiO_2$, propene $(C_3H_6)$ and isopropanol $(C_3H_8O)$ at temperatures below \SI{400}{\degreeCelsius} \cite{fictorie1994kinetic}. The chemical pathway is elucidated in Eq. \ref{propene_isopropanol}. These findings were subsequently corroborated by Fourier-transform infrared spectroscopy (FTIR) in a separate PECVD experiment conducted at 1 torr \cite{ahn2003kinetic}. It is possible that a similar chemical pathway provided the source of the three aforementioned elements on the samples.
Prior research in the literature during dusty plasma particle growth has utilized non-invasive characterization methods like atomic mass spectrometry and FTIR to identify molecules and nucleation within nanodusty plasmas \cite{kroesen1996situ, deschenaux1999investigations, ouaras2014situ, denysenko2018modeling,pattyn2018}. Although these techniques fall beyond the scope of the current study, they offer promising avenues for future investigations into the novel Ar/TTIP dusty plasma. 



\begin{align}
    Ti\left(OC_3H_7\right)_4 \rightarrow TiO_2 + 2C_3H_6 + 2C_3H_8O
    \label{propene_isopropanol}
\end{align}

To explain the potential presence and oxidation of C from the surface of the nanodust, it is essential to delve into the three key stages of plasma particle growth: nucleation, coagulation, and agglomeration, as extensively discussed in prior studies \cite{boufendi2011, kovacevic2012, stefanovic03, Donders22, Fridman96}. In the initial nucleation stage, plasma species such as radicals and ions are generated through interactions between the background plasma and reactive gases. These species subsequently engage in chemical reactions, forming clusters with sizes $\sim$1 nm. In the coagulation phase, these clusters collide and bond, leading to the creation of larger clusters with dimensions $\sim$10 nm. Coagulation is dominated by chemisorption of ions and clusters. Nucleation and coagulation processes occur within the first few hundred milliseconds and seconds, respectively \cite{ravi2009}.  The agglomeration stage, which can persist for several tens of seconds, involves the continued interaction of radicals and ions from the plasma, facilitating surface growth on the clusters and resulting in dust particles reaching sizes in the range of several hundred nm. This is dominated by physisorption of ions and radicals on the clusters formed. MD simulation in literature has shown that in Ar/$SiH_4$ dusty plasma particle growth, the ratio of chemisorption to physisorption decreased when the size of the dusty particles increased \cite{shi2021growth}. Literature has also shown that smaller sized $C_2H_x$ radicals exhibit longer lifetime than metallic radicals, during which their charge undergoes fluctuations, thereby causing surface growth on the dust through coulombic interactions \cite{winter2004dust}. This could potentially elucidate the presence of higher concentrations of Ti and O beneath certain layers of C, which subsequently undergo oxidation during the annealing process.

It is not a surprise that the bonding changed in the $TiO_2$ nanoparticles between coagulation and agglomeration. A similar situation has been reported in literature whereby NEXAFS was used to investigate the size-dependent chemical and physical properties of carbonaceous dust formed in a dusty plasma within a CCP system \cite{kovacevic2012}. The findings indicated that dust particles comprised an $sp^2$-rich core with a 10 nm diameter, likely formed during coagulation, and an $sp^2$-poor mantle with a diameter in the several hundred nm range, likely formed during agglomeration. It is plausible that a similar transformation in bonding may occur during the growth of $TiO_2$ particles. There remain intriguing prospects for future research into the bonding characteristics of dusty $TiO_2$ nanospheres. For example, further studies via TEM could help to examine the elemental distribution as a function of radius within the particles.

In this letter, we have introduced a new growth technique for $TiO_2$ via a TTIP metal-organic precursor that produces a new kind of dusty plasma. Our experiments showed that $TiO_2$ particles can be grown to homogenous sizes of 230 nm within 70 s. SEM images revealed a linear growth rate in particle radius and a homogeneous size distribution. Moreover, the samples were annealed to either anatase or rutile. Our results show room for future studies in particle growth using metal-organic precursors, characterization of particle chemistry and microstructure, and dusty plasma studies of interactions of particles with magnetic fields during growth.

\section*{Acknowledgement}
We gratefully acknowledge the financial support received from the National Science Foundation EPSCoR program (OIA-2148653), US Department of Energy - Plasma Science Facility (SC-0019176), and the NSF Major Research Instrumentation grant (NSF-DMR-2018794), which made this work possible.

\section*{Declaration of competing interest}

None.

\section*{DATA AVAILABILITY}
The data that support the findings of this study are available from the corresponding author upon reasonable request.

\section*{References}
\bibliography{aaaarefs}

\begin{thebibliography}{57}%
\makeatletter
\providecommand \@ifxundefined [1]{%
 \@ifx{#1\undefined}
}%
\providecommand \@ifnum [1]{%
 \ifnum #1\expandafter \@firstoftwo
 \else \expandafter \@secondoftwo
 \fi
}%
\providecommand \@ifx [1]{%
 \ifx #1\expandafter \@firstoftwo
 \else \expandafter \@secondoftwo
 \fi
}%
\providecommand \natexlab [1]{#1}%
\providecommand \enquote  [1]{``#1''}%
\providecommand \bibnamefont  [1]{#1}%
\providecommand \bibfnamefont [1]{#1}%
\providecommand \citenamefont [1]{#1}%
\providecommand \href@noop [0]{\@secondoftwo}%
\providecommand \href [0]{\begingroup \@sanitize@url \@href}%
\providecommand \@href[1]{\@@startlink{#1}\@@href}%
\providecommand \@@href[1]{\endgroup#1\@@endlink}%
\providecommand \@sanitize@url [0]{\catcode `\\12\catcode `\$12\catcode `\&12\catcode `\#12\catcode `\^12\catcode `\_12\catcode `\%12\relax}%
\providecommand \@@startlink[1]{}%
\providecommand \@@endlink[0]{}%
\providecommand \url  [0]{\begingroup\@sanitize@url \@url }%
\providecommand \@url [1]{\endgroup\@href {#1}{\urlprefix }}%
\providecommand \urlprefix  [0]{URL }%
\providecommand \Eprint [0]{\href }%
\providecommand \doibase [0]{http://dx.doi.org/}%
\providecommand \selectlanguage [0]{\@gobble}%
\providecommand \bibinfo  [0]{\@secondoftwo}%
\providecommand \bibfield  [0]{\@secondoftwo}%
\providecommand \translation [1]{[#1]}%
\providecommand \BibitemOpen [0]{}%
\providecommand \bibitemStop [0]{}%
\providecommand \bibitemNoStop [0]{.\EOS\space}%
\providecommand \EOS [0]{\spacefactor3000\relax}%
\providecommand \BibitemShut  [1]{\csname bibitem#1\endcsname}%
\let\auto@bib@innerbib\@empty
\bibitem [{\citenamefont {Kovacevic}\ \emph {et~al.}(2003)\citenamefont {Kovacevic}, \citenamefont {Stefanovic}, \citenamefont {Berndt},\ and\ \citenamefont {Winter}}]{kovavcevic2003}%
  \BibitemOpen
  \bibfield  {author} {\bibinfo {author} {\bibfnamefont {E.}~\bibnamefont {Kovacevic}}, \bibinfo {author} {\bibfnamefont {I.}~\bibnamefont {Stefanovic}}, \bibinfo {author} {\bibfnamefont {J.}~\bibnamefont {Berndt}}, \ and\ \bibinfo {author} {\bibfnamefont {J.}~\bibnamefont {Winter}},\ }\bibfield  {title} {\enquote {\bibinfo {title} {Infrared fingerprints and periodic formation of nanoparticles in {Ar}/{C$_2$H$_2$} plasmas},}\ }\href@noop {} {\bibfield  {journal} {\bibinfo  {journal} {J. Appl. Phys.}\ }\textbf {\bibinfo {volume} {93}},\ \bibinfo {pages} {2924--2930} (\bibinfo {year} {2003})}\BibitemShut {NoStop}%
\bibitem [{\citenamefont {Kovacevic}\ \emph {et~al.}(2009)\citenamefont {Kovacevic}, \citenamefont {Berndt}, \citenamefont {Stefanovic}, \citenamefont {Becker}, \citenamefont {Godde}, \citenamefont {Strunskus}, \citenamefont {Winter},\ and\ \citenamefont {Boufendi}}]{kovacevic2009}%
  \BibitemOpen
  \bibfield  {author} {\bibinfo {author} {\bibfnamefont {E.}~\bibnamefont {Kovacevic}}, \bibinfo {author} {\bibfnamefont {J.}~\bibnamefont {Berndt}}, \bibinfo {author} {\bibfnamefont {I.}~\bibnamefont {Stefanovic}}, \bibinfo {author} {\bibfnamefont {H.-W.}\ \bibnamefont {Becker}}, \bibinfo {author} {\bibfnamefont {C.}~\bibnamefont {Godde}}, \bibinfo {author} {\bibfnamefont {T.}~\bibnamefont {Strunskus}}, \bibinfo {author} {\bibfnamefont {J.}~\bibnamefont {Winter}}, \ and\ \bibinfo {author} {\bibfnamefont {L.}~\bibnamefont {Boufendi}},\ }\bibfield  {title} {\enquote {\bibinfo {title} {Formation and material analysis of plasma polymerized carbon nitride nanoparticles},}\ }\href@noop {} {\bibfield  {journal} {\bibinfo  {journal} {J. Appl. Phys.}\ }\textbf {\bibinfo {volume} {105}} (\bibinfo {year} {2009})}\BibitemShut {NoStop}%
\bibitem [{\citenamefont {Kovacevic}\ \emph {et~al.}(2012)\citenamefont {Kovacevic}, \citenamefont {Berndt}, \citenamefont {Strunskus},\ and\ \citenamefont {Boufendi}}]{kovacevic2012}%
  \BibitemOpen
  \bibfield  {author} {\bibinfo {author} {\bibfnamefont {E.}~\bibnamefont {Kovacevic}}, \bibinfo {author} {\bibfnamefont {J.}~\bibnamefont {Berndt}}, \bibinfo {author} {\bibfnamefont {T.}~\bibnamefont {Strunskus}}, \ and\ \bibinfo {author} {\bibfnamefont {L.}~\bibnamefont {Boufendi}},\ }\bibfield  {title} {\enquote {\bibinfo {title} {Size dependent characteristics of plasma synthesized carbonaceous nanoparticles},}\ }\href@noop {} {\bibfield  {journal} {\bibinfo  {journal} {J. Appl. Phys.}\ }\textbf {\bibinfo {volume} {112}} (\bibinfo {year} {2012})}\BibitemShut {NoStop}%
\bibitem [{\citenamefont {Couedel}\ \emph {et~al.}(2019)\citenamefont {Couedel}, \citenamefont {Artis}, \citenamefont {Khanal}, \citenamefont {Pardanaud}, \citenamefont {Coussan}, \citenamefont {LeBlanc}, \citenamefont {Hall}, \citenamefont {Jr.}, \citenamefont {Konopka}, \citenamefont {Park},\ and\ \citenamefont {Arnas}}]{couedel2019}%
  \BibitemOpen
  \bibfield  {author} {\bibinfo {author} {\bibfnamefont {L.}~\bibnamefont {Couedel}}, \bibinfo {author} {\bibfnamefont {D.}~\bibnamefont {Artis}}, \bibinfo {author} {\bibfnamefont {M.}~\bibnamefont {Khanal}}, \bibinfo {author} {\bibfnamefont {C.}~\bibnamefont {Pardanaud}}, \bibinfo {author} {\bibfnamefont {S.}~\bibnamefont {Coussan}}, \bibinfo {author} {\bibfnamefont {S.}~\bibnamefont {LeBlanc}}, \bibinfo {author} {\bibfnamefont {T.}~\bibnamefont {Hall}}, \bibinfo {author} {\bibfnamefont {E.~T.}\ \bibnamefont {Jr.}}, \bibinfo {author} {\bibfnamefont {U.}~\bibnamefont {Konopka}}, \bibinfo {author} {\bibfnamefont {M.}~\bibnamefont {Park}}, \ and\ \bibinfo {author} {\bibfnamefont {C.}~\bibnamefont {Arnas}},\ }\bibfield  {title} {\enquote {\bibinfo {title} {Influence of magnetic field strength on nanoparticle growth in a capacitively-coupled radio-frequency {Ar}/{C$_2$H$_2$} discharge},}\ }\href@noop {} {\bibfield  {journal} {\bibinfo  {journal} {Plasma Res. Express}\ }\textbf {\bibinfo {volume} {1}},\ \bibinfo
  {pages} {015012} (\bibinfo {year} {2019})}\BibitemShut {NoStop}%
\bibitem [{\citenamefont {Bouchoule}\ and\ \citenamefont {Boufendi}(1993)}]{bouchoule1993particulate}%
  \BibitemOpen
  \bibfield  {author} {\bibinfo {author} {\bibfnamefont {A.}~\bibnamefont {Bouchoule}}\ and\ \bibinfo {author} {\bibfnamefont {L.}~\bibnamefont {Boufendi}},\ }\bibfield  {title} {\enquote {\bibinfo {title} {Particulate formation and dusty plasma behaviour in argon-silane {RF} discharge},}\ }\href@noop {} {\bibfield  {journal} {\bibinfo  {journal} {Plasma Sources Sci. Technol.}\ }\textbf {\bibinfo {volume} {2}},\ \bibinfo {pages} {204} (\bibinfo {year} {1993})}\BibitemShut {NoStop}%
\bibitem [{\citenamefont {Watanabe}\ and\ \citenamefont {Shiratani}(1993)}]{watanabe1993growth}%
  \BibitemOpen
  \bibfield  {author} {\bibinfo {author} {\bibfnamefont {Y.~W.~Y.}\ \bibnamefont {Watanabe}}\ and\ \bibinfo {author} {\bibfnamefont {M.~S.~M.}\ \bibnamefont {Shiratani}},\ }\bibfield  {title} {\enquote {\bibinfo {title} {Growth kinetics and behavior of dust particles in silane plasmas},}\ }\href@noop {} {\bibfield  {journal} {\bibinfo  {journal} {Jpn. J. Appl. Phys.}\ }\textbf {\bibinfo {volume} {32}},\ \bibinfo {pages} {3074} (\bibinfo {year} {1993})}\BibitemShut {NoStop}%
\bibitem [{\citenamefont {Boufendi}\ and\ \citenamefont {Bouchoule}(1994)}]{boufendi1994}%
  \BibitemOpen
  \bibfield  {author} {\bibinfo {author} {\bibfnamefont {L.}~\bibnamefont {Boufendi}}\ and\ \bibinfo {author} {\bibfnamefont {A.}~\bibnamefont {Bouchoule}},\ }\bibfield  {title} {\enquote {\bibinfo {title} {Particle nucleation and growth in a low-pressure argon-silane discharge},}\ }\href@noop {} {\bibfield  {journal} {\bibinfo  {journal} {Plasma Sources Sci. Technol.}\ }\textbf {\bibinfo {volume} {3}},\ \bibinfo {pages} {262} (\bibinfo {year} {1994})}\BibitemShut {NoStop}%
\bibitem [{\citenamefont {Ganguly}\ \emph {et~al.}(1993)\citenamefont {Ganguly}, \citenamefont {Garscadden}, \citenamefont {Williams},\ and\ \citenamefont {Haaland}}]{ganguly1993growth}%
  \BibitemOpen
  \bibfield  {author} {\bibinfo {author} {\bibfnamefont {B.}~\bibnamefont {Ganguly}}, \bibinfo {author} {\bibfnamefont {A.}~\bibnamefont {Garscadden}}, \bibinfo {author} {\bibfnamefont {J.}~\bibnamefont {Williams}}, \ and\ \bibinfo {author} {\bibfnamefont {P.}~\bibnamefont {Haaland}},\ }\bibfield  {title} {\enquote {\bibinfo {title} {Growth and morphology of carbon grains},}\ }\href@noop {} {\bibfield  {journal} {\bibinfo  {journal} {J. Vac. Sci. Technol., A}\ }\textbf {\bibinfo {volume} {11}},\ \bibinfo {pages} {1119--1125} (\bibinfo {year} {1993})}\BibitemShut {NoStop}%
\bibitem [{\citenamefont {Garscadden}\ \emph {et~al.}(1994)\citenamefont {Garscadden}, \citenamefont {Ganguly}, \citenamefont {Haaland},\ and\ \citenamefont {Williams}}]{garscadden1994overview}%
  \BibitemOpen
  \bibfield  {author} {\bibinfo {author} {\bibfnamefont {A.}~\bibnamefont {Garscadden}}, \bibinfo {author} {\bibfnamefont {B.}~\bibnamefont {Ganguly}}, \bibinfo {author} {\bibfnamefont {P.}~\bibnamefont {Haaland}}, \ and\ \bibinfo {author} {\bibfnamefont {J.}~\bibnamefont {Williams}},\ }\bibfield  {title} {\enquote {\bibinfo {title} {Overview of growth and behaviour of clusters and particles in plasmas},}\ }\href@noop {} {\bibfield  {journal} {\bibinfo  {journal} {Plasma Sources Sci. Technol.}\ }\textbf {\bibinfo {volume} {3}},\ \bibinfo {pages} {239} (\bibinfo {year} {1994})}\BibitemShut {NoStop}%
\bibitem [{\citenamefont {Cao}\ and\ \citenamefont {Matsoukas}(2002)}]{cao2002deposition}%
  \BibitemOpen
  \bibfield  {author} {\bibinfo {author} {\bibfnamefont {J.}~\bibnamefont {Cao}}\ and\ \bibinfo {author} {\bibfnamefont {T.}~\bibnamefont {Matsoukas}},\ }\bibfield  {title} {\enquote {\bibinfo {title} {Deposition kinetics on particles in a dusty plasma reactor},}\ }\href@noop {} {\bibfield  {journal} {\bibinfo  {journal} {J. Appl. Phys.}\ }\textbf {\bibinfo {volume} {92}},\ \bibinfo {pages} {2916--2922} (\bibinfo {year} {2002})}\BibitemShut {NoStop}%
\bibitem [{\citenamefont {Galli}\ and\ \citenamefont {Kortshagen}(2009)}]{galli2009charging}%
  \BibitemOpen
  \bibfield  {author} {\bibinfo {author} {\bibfnamefont {F.}~\bibnamefont {Galli}}\ and\ \bibinfo {author} {\bibfnamefont {U.~R.}\ \bibnamefont {Kortshagen}},\ }\bibfield  {title} {\enquote {\bibinfo {title} {Charging, coagulation, and heating model of nanoparticles in a low-pressure plasma accounting for ion--neutral collisions},}\ }\href@noop {} {\bibfield  {journal} {\bibinfo  {journal} {IEEE Trans. Plasma Sci.}\ }\textbf {\bibinfo {volume} {38}},\ \bibinfo {pages} {803--809} (\bibinfo {year} {2009})}\BibitemShut {NoStop}%
\bibitem [{\citenamefont {Chutia}\ \emph {et~al.}(2021)\citenamefont {Chutia}, \citenamefont {Deka}, \citenamefont {Bailung}, \citenamefont {Sharma},\ and\ \citenamefont {Bailung}}]{chutia2021nanodusty}%
  \BibitemOpen
  \bibfield  {author} {\bibinfo {author} {\bibfnamefont {B.}~\bibnamefont {Chutia}}, \bibinfo {author} {\bibfnamefont {T.}~\bibnamefont {Deka}}, \bibinfo {author} {\bibfnamefont {Y.}~\bibnamefont {Bailung}}, \bibinfo {author} {\bibfnamefont {S.}~\bibnamefont {Sharma}}, \ and\ \bibinfo {author} {\bibfnamefont {H.}~\bibnamefont {Bailung}},\ }\bibfield  {title} {\enquote {\bibinfo {title} {A nanodusty plasma experiment to create extended dust clouds using reactive argon acetylene plasmas},}\ }\href@noop {} {\bibfield  {journal} {\bibinfo  {journal} {Phys. Plasmas}\ }\textbf {\bibinfo {volume} {28}} (\bibinfo {year} {2021})}\BibitemShut {NoStop}%
\bibitem [{\citenamefont {Kortshagen}(2009)}]{kortshagen2009nonthermal}%
  \BibitemOpen
  \bibfield  {author} {\bibinfo {author} {\bibfnamefont {U.}~\bibnamefont {Kortshagen}},\ }\bibfield  {title} {\enquote {\bibinfo {title} {Nonthermal plasma synthesis of semiconductor nanocrystals},}\ }\href@noop {} {\bibfield  {journal} {\bibinfo  {journal} {Journal of Physics D: Applied Physics}\ }\textbf {\bibinfo {volume} {42}},\ \bibinfo {pages} {113001} (\bibinfo {year} {2009})}\BibitemShut {NoStop}%
\bibitem [{\citenamefont {Boufendi}\ \emph {et~al.}(2011)\citenamefont {Boufendi}, \citenamefont {Jouanny}, \citenamefont {Kovacevic}, \citenamefont {Berndt},\ and\ \citenamefont {Mikikian}}]{boufendi2011}%
  \BibitemOpen
  \bibfield  {author} {\bibinfo {author} {\bibfnamefont {L.}~\bibnamefont {Boufendi}}, \bibinfo {author} {\bibfnamefont {M.}~\bibnamefont {Jouanny}}, \bibinfo {author} {\bibfnamefont {E.}~\bibnamefont {Kovacevic}}, \bibinfo {author} {\bibfnamefont {J.}~\bibnamefont {Berndt}}, \ and\ \bibinfo {author} {\bibfnamefont {M.}~\bibnamefont {Mikikian}},\ }\bibfield  {title} {\enquote {\bibinfo {title} {Dusty plasma for nanotechnology},}\ }\href@noop {} {\bibfield  {journal} {\bibinfo  {journal} {J. Phys. D: Appl. Phys.}\ }\textbf {\bibinfo {volume} {44}},\ \bibinfo {pages} {174035} (\bibinfo {year} {2011})}\BibitemShut {NoStop}%
\bibitem [{\citenamefont {Despax}, \citenamefont {Makasheva},\ and\ \citenamefont {Caquineau}(2012)}]{despax2012}%
  \BibitemOpen
  \bibfield  {author} {\bibinfo {author} {\bibfnamefont {B.}~\bibnamefont {Despax}}, \bibinfo {author} {\bibfnamefont {K.}~\bibnamefont {Makasheva}}, \ and\ \bibinfo {author} {\bibfnamefont {H.}~\bibnamefont {Caquineau}},\ }\bibfield  {title} {\enquote {\bibinfo {title} {Cyclic powder formation during pulsed injection of hexamethyldisiloxane in an axially asymmetric radiofrequency argon discharge},}\ }\href@noop {} {\bibfield  {journal} {\bibinfo  {journal} {J. Appl. Phys.}\ }\textbf {\bibinfo {volume} {112}} (\bibinfo {year} {2012})}\BibitemShut {NoStop}%
\bibitem [{\citenamefont {Pattyn}\ \emph {et~al.}(2018)\citenamefont {Pattyn}, \citenamefont {Kovacevic}, \citenamefont {Hussain}, \citenamefont {Dias}, \citenamefont {Lecas},\ and\ \citenamefont {Berndt}}]{pattyn2018}%
  \BibitemOpen
  \bibfield  {author} {\bibinfo {author} {\bibfnamefont {C.}~\bibnamefont {Pattyn}}, \bibinfo {author} {\bibfnamefont {E.}~\bibnamefont {Kovacevic}}, \bibinfo {author} {\bibfnamefont {S.}~\bibnamefont {Hussain}}, \bibinfo {author} {\bibfnamefont {A.}~\bibnamefont {Dias}}, \bibinfo {author} {\bibfnamefont {T.}~\bibnamefont {Lecas}}, \ and\ \bibinfo {author} {\bibfnamefont {J.}~\bibnamefont {Berndt}},\ }\bibfield  {title} {\enquote {\bibinfo {title} {Nanoparticle formation in a low pressure argon/aniline {RF} plasma},}\ }\href@noop {} {\bibfield  {journal} {\bibinfo  {journal} {Appl. Phys. Lett.}\ }\textbf {\bibinfo {volume} {112}} (\bibinfo {year} {2018})}\BibitemShut {NoStop}%
\bibitem [{\citenamefont {Cameron}\ \emph {et~al.}(2023)\citenamefont {Cameron}, \citenamefont {Klause}, \citenamefont {Andaraarachchi}, \citenamefont {Xiong}, \citenamefont {Reed}, \citenamefont {Thapa}, \citenamefont {Wu},\ and\ \citenamefont {Kortshagen}}]{cameron2023capacitively}%
  \BibitemOpen
  \bibfield  {author} {\bibinfo {author} {\bibfnamefont {T.~J.}\ \bibnamefont {Cameron}}, \bibinfo {author} {\bibfnamefont {B.}~\bibnamefont {Klause}}, \bibinfo {author} {\bibfnamefont {H.~P.}\ \bibnamefont {Andaraarachchi}}, \bibinfo {author} {\bibfnamefont {Z.}~\bibnamefont {Xiong}}, \bibinfo {author} {\bibfnamefont {C.}~\bibnamefont {Reed}}, \bibinfo {author} {\bibfnamefont {D.}~\bibnamefont {Thapa}}, \bibinfo {author} {\bibfnamefont {C.-C.}\ \bibnamefont {Wu}}, \ and\ \bibinfo {author} {\bibfnamefont {U.~R.}\ \bibnamefont {Kortshagen}},\ }\bibfield  {title} {\enquote {\bibinfo {title} {Capacitively coupled nonthermal plasma synthesis of aluminum nanocrystals for enhanced yield and size control},}\ }\href@noop {} {\bibfield  {journal} {\bibinfo  {journal} {Nanotechnology}\ } (\bibinfo {year} {2023})}\BibitemShut {NoStop}%
\bibitem [{\citenamefont {Wang}, \citenamefont {Lin},\ and\ \citenamefont {Hon}(1997)}]{wang1997dependence}%
  \BibitemOpen
  \bibfield  {author} {\bibinfo {author} {\bibfnamefont {H.}~\bibnamefont {Wang}}, \bibinfo {author} {\bibfnamefont {C.}~\bibnamefont {Lin}}, \ and\ \bibinfo {author} {\bibfnamefont {M.}~\bibnamefont {Hon}},\ }\bibfield  {title} {\enquote {\bibinfo {title} {The dependence of hardness on the density of amorphous alumina thin films by pecvd},}\ }\href@noop {} {\bibfield  {journal} {\bibinfo  {journal} {Thin Solid Films}\ }\textbf {\bibinfo {volume} {310}},\ \bibinfo {pages} {260--264} (\bibinfo {year} {1997})}\BibitemShut {NoStop}%
\bibitem [{\citenamefont {Shirafuji}\ \emph {et~al.}(1999)\citenamefont {Shirafuji}, \citenamefont {Miyazaki}, \citenamefont {Nakagami}, \citenamefont {Hayashi},\ and\ \citenamefont {Nishino}}]{Shirafuji99}%
  \BibitemOpen
  \bibfield  {author} {\bibinfo {author} {\bibfnamefont {T.}~\bibnamefont {Shirafuji}}, \bibinfo {author} {\bibfnamefont {Y.}~\bibnamefont {Miyazaki}}, \bibinfo {author} {\bibfnamefont {Y.}~\bibnamefont {Nakagami}}, \bibinfo {author} {\bibfnamefont {Y.}~\bibnamefont {Hayashi}}, \ and\ \bibinfo {author} {\bibfnamefont {S.}~\bibnamefont {Nishino}},\ }\bibfield  {title} {\enquote {\bibinfo {title} {Plasma copolymerization of tetrafluoroethylene/hexamethyldisiloxane and in situ fourier transform infrared spectroscopy of its gas phase},}\ }\href@noop {} {\bibfield  {journal} {\bibinfo  {journal} {Jpn. J. Appl. Phys.}\ }\textbf {\bibinfo {volume} {38}},\ \bibinfo {pages} {4520--4526} (\bibinfo {year} {1999})}\BibitemShut {NoStop}%
\bibitem [{\citenamefont {Shioya}\ \emph {et~al.}(2005)\citenamefont {Shioya}, \citenamefont {Shimoda}, \citenamefont {Maeda}, \citenamefont {Ohdaira}, \citenamefont {Suzuki},\ and\ \citenamefont {Seino}}]{Shioya05}%
  \BibitemOpen
  \bibfield  {author} {\bibinfo {author} {\bibfnamefont {Y.}~\bibnamefont {Shioya}}, \bibinfo {author} {\bibfnamefont {H.}~\bibnamefont {Shimoda}}, \bibinfo {author} {\bibfnamefont {K.}~\bibnamefont {Maeda}}, \bibinfo {author} {\bibfnamefont {T.}~\bibnamefont {Ohdaira}}, \bibinfo {author} {\bibfnamefont {R.}~\bibnamefont {Suzuki}}, \ and\ \bibinfo {author} {\bibfnamefont {Y.}~\bibnamefont {Seino}},\ }\bibfield  {title} {\enquote {\bibinfo {title} {Low-k sioch film deposited by plasma-enhanced chemical vapor deposition using hexamethyldisiloxane and water vapor},}\ }\href@noop {} {\bibfield  {journal} {\bibinfo  {journal} {Jpn. J. Appl. Phys.}\ }\textbf {\bibinfo {volume} {44}},\ \bibinfo {pages} {3879--3884} (\bibinfo {year} {2005})}\BibitemShut {NoStop}%
\bibitem [{\citenamefont {Airoudj}\ \emph {et~al.}(2008)\citenamefont {Airoudj}, \citenamefont {Debarnot}, \citenamefont {Beche},\ and\ \citenamefont {Poncin-Epaillard}}]{airoudj2008}%
  \BibitemOpen
  \bibfield  {author} {\bibinfo {author} {\bibfnamefont {A.}~\bibnamefont {Airoudj}}, \bibinfo {author} {\bibfnamefont {D.}~\bibnamefont {Debarnot}}, \bibinfo {author} {\bibfnamefont {B.}~\bibnamefont {Beche}}, \ and\ \bibinfo {author} {\bibfnamefont {F.}~\bibnamefont {Poncin-Epaillard}},\ }\bibfield  {title} {\enquote {\bibinfo {title} {New sensitive layer based on pulsed plasma-polymerized aniline for integrated optical ammonia sensor},}\ }\href@noop {} {\bibfield  {journal} {\bibinfo  {journal} {Anal. Chim. Acta}\ }\textbf {\bibinfo {volume} {626}},\ \bibinfo {pages} {44--52} (\bibinfo {year} {2008})}\BibitemShut {NoStop}%
\bibitem [{\citenamefont {Lee}\ \emph {et~al.}(1994)\citenamefont {Lee}, \citenamefont {Woo}, \citenamefont {Kim}, \citenamefont {Choi},\ and\ \citenamefont {Oh}}]{Lee94}%
  \BibitemOpen
  \bibfield  {author} {\bibinfo {author} {\bibfnamefont {W.}~\bibnamefont {Lee}}, \bibinfo {author} {\bibfnamefont {S.~I.}\ \bibnamefont {Woo}}, \bibinfo {author} {\bibfnamefont {J.}~\bibnamefont {Kim}}, \bibinfo {author} {\bibfnamefont {S.}~\bibnamefont {Choi}}, \ and\ \bibinfo {author} {\bibfnamefont {K.}~\bibnamefont {Oh}},\ }\bibfield  {title} {\enquote {\bibinfo {title} {Preparation and properties of amorphous {TiO}$_2$ thin films by plasma enhanced chemical vapor deposition},}\ }\href@noop {} {\bibfield  {journal} {\bibinfo  {journal} {Thin Solid Films}\ }\textbf {\bibinfo {volume} {237}},\ \bibinfo {pages} {105--111} (\bibinfo {year} {1994})}\BibitemShut {NoStop}%
\bibitem [{\citenamefont {Yang}\ and\ \citenamefont {Wolden}(2006)}]{Yang06}%
  \BibitemOpen
  \bibfield  {author} {\bibinfo {author} {\bibfnamefont {W.}~\bibnamefont {Yang}}\ and\ \bibinfo {author} {\bibfnamefont {C.}~\bibnamefont {Wolden}},\ }\bibfield  {title} {\enquote {\bibinfo {title} {Plasma-enhanced chemical vapor deposition of {TiO}$_2$ thin films for dielectric applications},}\ }\href@noop {} {\bibfield  {journal} {\bibinfo  {journal} {Thin Solid Films}\ }\textbf {\bibinfo {volume} {515}},\ \bibinfo {pages} {1708--1713} (\bibinfo {year} {2006})}\BibitemShut {NoStop}%
\bibitem [{\citenamefont {Aarik}\ \emph {et~al.}(2000)\citenamefont {Aarik}, \citenamefont {Aidla}, \citenamefont {Uustare}, \citenamefont {Ritala},\ and\ \citenamefont {Leskel{\"a}}}]{aarik2000titanium}%
  \BibitemOpen
  \bibfield  {author} {\bibinfo {author} {\bibfnamefont {J.}~\bibnamefont {Aarik}}, \bibinfo {author} {\bibfnamefont {A.}~\bibnamefont {Aidla}}, \bibinfo {author} {\bibfnamefont {T.}~\bibnamefont {Uustare}}, \bibinfo {author} {\bibfnamefont {M.}~\bibnamefont {Ritala}}, \ and\ \bibinfo {author} {\bibfnamefont {M.}~\bibnamefont {Leskel{\"a}}},\ }\bibfield  {title} {\enquote {\bibinfo {title} {Titanium isopropoxide as a precursor for atomic layer deposition: characterization of titanium dioxide growth process},}\ }\href@noop {} {\bibfield  {journal} {\bibinfo  {journal} {Appl. Surf. Sci.}\ }\textbf {\bibinfo {volume} {161}},\ \bibinfo {pages} {385--395} (\bibinfo {year} {2000})}\BibitemShut {NoStop}%
\bibitem [{\citenamefont {Lee}\ \emph {et~al.}(2013)\citenamefont {Lee}, \citenamefont {Lee}, \citenamefont {Han}, \citenamefont {Jeon}, \citenamefont {Park}, \citenamefont {Jang}, \citenamefont {Yoon},\ and\ \citenamefont {Jeon}}]{lee2013deposition}%
  \BibitemOpen
  \bibfield  {author} {\bibinfo {author} {\bibfnamefont {J.}~\bibnamefont {Lee}}, \bibinfo {author} {\bibfnamefont {S.~J.}\ \bibnamefont {Lee}}, \bibinfo {author} {\bibfnamefont {W.~B.}\ \bibnamefont {Han}}, \bibinfo {author} {\bibfnamefont {H.}~\bibnamefont {Jeon}}, \bibinfo {author} {\bibfnamefont {J.}~\bibnamefont {Park}}, \bibinfo {author} {\bibfnamefont {W.}~\bibnamefont {Jang}}, \bibinfo {author} {\bibfnamefont {C.~S.}\ \bibnamefont {Yoon}}, \ and\ \bibinfo {author} {\bibfnamefont {H.}~\bibnamefont {Jeon}},\ }\bibfield  {title} {\enquote {\bibinfo {title} {Deposition temperature dependence of titanium oxide thin films grown by remote-plasma atomic layer deposition},}\ }\href@noop {} {\bibfield  {journal} {\bibinfo  {journal} {Phys. Status Solidi A}\ }\textbf {\bibinfo {volume} {210}},\ \bibinfo {pages} {276--284} (\bibinfo {year} {2013})}\BibitemShut {NoStop}%
\bibitem [{\citenamefont {Juma}\ \emph {et~al.}(2015)\citenamefont {Juma}, \citenamefont {Acik}, \citenamefont {Mikli}, \citenamefont {Mere},\ and\ \citenamefont {Krunks}}]{juma2015effect}%
  \BibitemOpen
  \bibfield  {author} {\bibinfo {author} {\bibfnamefont {A.~O.}\ \bibnamefont {Juma}}, \bibinfo {author} {\bibfnamefont {I.~O.}\ \bibnamefont {Acik}}, \bibinfo {author} {\bibfnamefont {V.}~\bibnamefont {Mikli}}, \bibinfo {author} {\bibfnamefont {A.}~\bibnamefont {Mere}}, \ and\ \bibinfo {author} {\bibfnamefont {M.}~\bibnamefont {Krunks}},\ }\bibfield  {title} {\enquote {\bibinfo {title} {Effect of solution composition on anatase to rutile transformation of sprayed {TiO}$_2$ thin films},}\ }\href@noop {} {\bibfield  {journal} {\bibinfo  {journal} {Thin Solid Films}\ }\textbf {\bibinfo {volume} {594}},\ \bibinfo {pages} {287--292} (\bibinfo {year} {2015})}\BibitemShut {NoStop}%
\bibitem [{\citenamefont {Muaz}\ \emph {et~al.}(2016)\citenamefont {Muaz}, \citenamefont {Hashim}, \citenamefont {Ibrahim}, \citenamefont {Thong}, \citenamefont {Mohktar},\ and\ \citenamefont {Liu}}]{muaz2016effect}%
  \BibitemOpen
  \bibfield  {author} {\bibinfo {author} {\bibfnamefont {A.}~\bibnamefont {Muaz}}, \bibinfo {author} {\bibfnamefont {U.}~\bibnamefont {Hashim}}, \bibinfo {author} {\bibfnamefont {F.}~\bibnamefont {Ibrahim}}, \bibinfo {author} {\bibfnamefont {K.}~\bibnamefont {Thong}}, \bibinfo {author} {\bibfnamefont {M.~S.}\ \bibnamefont {Mohktar}}, \ and\ \bibinfo {author} {\bibfnamefont {W.-W.}\ \bibnamefont {Liu}},\ }\bibfield  {title} {\enquote {\bibinfo {title} {Effect of annealing temperatures on the morphology, optical and electrical properties of tio 2 thin films synthesized by the sol--gel method and deposited on {Al}/{TiO}$_2$/{SiO}$_2$/p-{Si}},}\ }\href@noop {} {\bibfield  {journal} {\bibinfo  {journal} {Microsyst. Technol.}\ }\textbf {\bibinfo {volume} {22}},\ \bibinfo {pages} {871--881} (\bibinfo {year} {2016})}\BibitemShut {NoStop}%
\bibitem [{\citenamefont {Jalan}\ \emph {et~al.}(2009{\natexlab{a}})\citenamefont {Jalan}, \citenamefont {Engel-Herbert}, \citenamefont {Cagnon},\ and\ \citenamefont {Stemmer}}]{jalanTiO2}%
  \BibitemOpen
  \bibfield  {author} {\bibinfo {author} {\bibfnamefont {B.}~\bibnamefont {Jalan}}, \bibinfo {author} {\bibfnamefont {R.}~\bibnamefont {Engel-Herbert}}, \bibinfo {author} {\bibfnamefont {J.}~\bibnamefont {Cagnon}}, \ and\ \bibinfo {author} {\bibfnamefont {S.}~\bibnamefont {Stemmer}},\ }\bibfield  {title} {\enquote {\bibinfo {title} {{Growth modes in metal-organic molecular beam epitaxy of {TiO}$_2$ on r-plane sapphire}},}\ }\href@noop {} {\bibfield  {journal} {\bibinfo  {journal} {J. Vac. Sci. Technol. A}\ }\textbf {\bibinfo {volume} {27}},\ \bibinfo {pages} {230--233} (\bibinfo {year} {2009}{\natexlab{a}})}\BibitemShut {NoStop}%
\bibitem [{\citenamefont {Jalan}\ \emph {et~al.}(2009{\natexlab{b}})\citenamefont {Jalan}, \citenamefont {Engel-Herbert}, \citenamefont {Wright},\ and\ \citenamefont {Stemmer}}]{jalanSTO}%
  \BibitemOpen
  \bibfield  {author} {\bibinfo {author} {\bibfnamefont {B.}~\bibnamefont {Jalan}}, \bibinfo {author} {\bibfnamefont {R.}~\bibnamefont {Engel-Herbert}}, \bibinfo {author} {\bibfnamefont {N.~J.}\ \bibnamefont {Wright}}, \ and\ \bibinfo {author} {\bibfnamefont {S.}~\bibnamefont {Stemmer}},\ }\bibfield  {title} {\enquote {\bibinfo {title} {{Growth of high-quality {SrTiO}$_3$ films using a hybrid molecular beam epitaxy approach}},}\ }\href@noop {} {\bibfield  {journal} {\bibinfo  {journal} {J. Vac. Sci. Technol. A}\ }\textbf {\bibinfo {volume} {27}},\ \bibinfo {pages} {461--464} (\bibinfo {year} {2009}{\natexlab{b}})}\BibitemShut {NoStop}%
\bibitem [{\citenamefont {Thapa}\ \emph {et~al.}(2021)\citenamefont {Thapa}, \citenamefont {Provence}, \citenamefont {Jessup}, \citenamefont {Lapano}, \citenamefont {Brahlek}, \citenamefont {Sadowski}, \citenamefont {Reinke}, \citenamefont {Jin},\ and\ \citenamefont {Comes}}]{thapaSTO}%
  \BibitemOpen
  \bibfield  {author} {\bibinfo {author} {\bibfnamefont {S.}~\bibnamefont {Thapa}}, \bibinfo {author} {\bibfnamefont {S.~R.}\ \bibnamefont {Provence}}, \bibinfo {author} {\bibfnamefont {D.}~\bibnamefont {Jessup}}, \bibinfo {author} {\bibfnamefont {J.}~\bibnamefont {Lapano}}, \bibinfo {author} {\bibfnamefont {M.}~\bibnamefont {Brahlek}}, \bibinfo {author} {\bibfnamefont {J.~T.}\ \bibnamefont {Sadowski}}, \bibinfo {author} {\bibfnamefont {P.}~\bibnamefont {Reinke}}, \bibinfo {author} {\bibfnamefont {W.}~\bibnamefont {Jin}}, \ and\ \bibinfo {author} {\bibfnamefont {R.~B.}\ \bibnamefont {Comes}},\ }\bibfield  {title} {\enquote {\bibinfo {title} {{Correlating surface stoichiometry and termination in {SrTiO}$_3$ films grown by hybrid molecular beam epitaxy}},}\ }\href@noop {} {\bibfield  {journal} {\bibinfo  {journal} {J. Vac. Sci. Technol. A}\ }\textbf {\bibinfo {volume} {39}},\ \bibinfo {pages} {053203} (\bibinfo {year} {2021})}\BibitemShut {NoStop}%
\bibitem [{\citenamefont {Lapano}\ \emph {et~al.}(2019)\citenamefont {Lapano}, \citenamefont {Brahlek}, \citenamefont {Zhang}, \citenamefont {Roth}, \citenamefont {Pogrebnyakov},\ and\ \citenamefont {Engel-Herbert}}]{engel-herbertSTO}%
  \BibitemOpen
  \bibfield  {author} {\bibinfo {author} {\bibfnamefont {J.}~\bibnamefont {Lapano}}, \bibinfo {author} {\bibfnamefont {M.}~\bibnamefont {Brahlek}}, \bibinfo {author} {\bibfnamefont {L.}~\bibnamefont {Zhang}}, \bibinfo {author} {\bibfnamefont {J.}~\bibnamefont {Roth}}, \bibinfo {author} {\bibfnamefont {A.}~\bibnamefont {Pogrebnyakov}}, \ and\ \bibinfo {author} {\bibfnamefont {R.}~\bibnamefont {Engel-Herbert}},\ }\bibfield  {title} {\enquote {\bibinfo {title} {{Scaling growth rates for perovskite oxide virtual substrates on silicon}},}\ }\href {\doibase 10.1038/s41467-019-10273-2} {\bibfield  {journal} {\bibinfo  {journal} {Nat. Commun.}\ }\textbf {\bibinfo {volume} {10}},\ \bibinfo {pages} {2464} (\bibinfo {year} {2019})}\BibitemShut {NoStop}%
\bibitem [{\citenamefont {Huang}\ and\ \citenamefont {Chen}(2002)}]{huang2002comparison}%
  \BibitemOpen
  \bibfield  {author} {\bibinfo {author} {\bibfnamefont {S.}~\bibnamefont {Huang}}\ and\ \bibinfo {author} {\bibfnamefont {J.-S.}\ \bibnamefont {Chen}},\ }\bibfield  {title} {\enquote {\bibinfo {title} {Comparison of the characteristics of {TiO}$_2$ films prepared by low-pressure and plasma-enhanced chemical vapor deposition},}\ }\href@noop {} {\bibfield  {journal} {\bibinfo  {journal} {J. Mater. Sci.: Mater. Electron.}\ }\textbf {\bibinfo {volume} {13}},\ \bibinfo {pages} {77--81} (\bibinfo {year} {2002})}\BibitemShut {NoStop}%
\bibitem [{\citenamefont {Nguyen}\ \emph {et~al.}(2013)\citenamefont {Nguyen}, \citenamefont {Kim}, \citenamefont {Park},\ and\ \citenamefont {Kim}}]{Nguyen13}%
  \BibitemOpen
  \bibfield  {author} {\bibinfo {author} {\bibfnamefont {H.}~\bibnamefont {Nguyen}}, \bibinfo {author} {\bibfnamefont {D.}~\bibnamefont {Kim}}, \bibinfo {author} {\bibfnamefont {D.}~\bibnamefont {Park}}, \ and\ \bibinfo {author} {\bibfnamefont {K.}~\bibnamefont {Kim}},\ }\bibfield  {title} {\enquote {\bibinfo {title} {Effect of initial precursor concentration on {TiO}$_2$ thin film nanostructures prepared by {PCVD} system},}\ }\href@noop {} {\bibfield  {journal} {\bibinfo  {journal} {J. Energy Chem.}\ }\textbf {\bibinfo {volume} {22}},\ \bibinfo {pages} {375--381} (\bibinfo {year} {2013})}\BibitemShut {NoStop}%
\bibitem [{\citenamefont {Stoner}\ \emph {et~al.}(1992)\citenamefont {Stoner}, \citenamefont {Ma}, \citenamefont {Wolter},\ and\ \citenamefont {Glass}}]{Stoner92}%
  \BibitemOpen
  \bibfield  {author} {\bibinfo {author} {\bibfnamefont {B.}~\bibnamefont {Stoner}}, \bibinfo {author} {\bibfnamefont {G.-H.~M.}\ \bibnamefont {Ma}}, \bibinfo {author} {\bibfnamefont {S.~D.}\ \bibnamefont {Wolter}}, \ and\ \bibinfo {author} {\bibfnamefont {J.}~\bibnamefont {Glass}},\ }\bibfield  {title} {\enquote {\bibinfo {title} {Characterization of bias-enhanced nucleation of diamond on silicon by in vacuo surface analysis and transmission electron microscopy},}\ }\href@noop {} {\bibfield  {journal} {\bibinfo  {journal} {Phys. Rev. B}\ }\textbf {\bibinfo {volume} {45}},\ \bibinfo {pages} {11 067--11 084} (\bibinfo {year} {1992})}\BibitemShut {NoStop}%
\bibitem [{\citenamefont {Kuhr}, \citenamefont {Reinke},\ and\ \citenamefont {Kulisch}(1995)}]{Kuhr95}%
  \BibitemOpen
  \bibfield  {author} {\bibinfo {author} {\bibfnamefont {M.}~\bibnamefont {Kuhr}}, \bibinfo {author} {\bibfnamefont {S.}~\bibnamefont {Reinke}}, \ and\ \bibinfo {author} {\bibfnamefont {W.}~\bibnamefont {Kulisch}},\ }\bibfield  {title} {\enquote {\bibinfo {title} {Nucleation of cubic boron nitride (c-{BN}) with ion-induced plasma-enhanced {CVD}},}\ }\href@noop {} {\bibfield  {journal} {\bibinfo  {journal} {Diamond Relat. Mater.}\ }\textbf {\bibinfo {volume} {4}},\ \bibinfo {pages} {375--380} (\bibinfo {year} {1995})}\BibitemShut {NoStop}%
\bibitem [{\citenamefont {Ramkorun}, \citenamefont {Chakrabarty},\ and\ \citenamefont {Catledge}(2021)}]{Ramkorun21}%
  \BibitemOpen
  \bibfield  {author} {\bibinfo {author} {\bibfnamefont {B.}~\bibnamefont {Ramkorun}}, \bibinfo {author} {\bibfnamefont {K.}~\bibnamefont {Chakrabarty}}, \ and\ \bibinfo {author} {\bibfnamefont {S.}~\bibnamefont {Catledge}},\ }\bibfield  {title} {\enquote {\bibinfo {title} {Effects of direct current bias on nucleation density of superhard boron-rich boron carbide films made by microwave plasma chemical vapor deposition},}\ }\href@noop {} {\bibfield  {journal} {\bibinfo  {journal} {Mater. Res. Express}\ }\textbf {\bibinfo {volume} {8}},\ \bibinfo {pages} {1--12} (\bibinfo {year} {2021})}\BibitemShut {NoStop}%
\bibitem [{\citenamefont {Ravi}\ and\ \citenamefont {Girshick}(2009)}]{ravi2009}%
  \BibitemOpen
  \bibfield  {author} {\bibinfo {author} {\bibfnamefont {L.}~\bibnamefont {Ravi}}\ and\ \bibinfo {author} {\bibfnamefont {S.}~\bibnamefont {Girshick}},\ }\bibfield  {title} {\enquote {\bibinfo {title} {Coagulation of nanoparticles in a plasma},}\ }\href@noop {} {\bibfield  {journal} {\bibinfo  {journal} {Phys. Rev. E}\ }\textbf {\bibinfo {volume} {79}},\ \bibinfo {pages} {026408} (\bibinfo {year} {2009})}\BibitemShut {NoStop}%
\bibitem [{\citenamefont {Groth}\ \emph {et~al.}(2015)\citenamefont {Groth}, \citenamefont {Greiner}, \citenamefont {Tadsen},\ and\ \citenamefont {Piel}}]{groth2015}%
  \BibitemOpen
  \bibfield  {author} {\bibinfo {author} {\bibfnamefont {S.}~\bibnamefont {Groth}}, \bibinfo {author} {\bibfnamefont {F.}~\bibnamefont {Greiner}}, \bibinfo {author} {\bibfnamefont {B.}~\bibnamefont {Tadsen}}, \ and\ \bibinfo {author} {\bibfnamefont {A.}~\bibnamefont {Piel}},\ }\bibfield  {title} {\enquote {\bibinfo {title} {Kinetic mie ellipsometry to determine the time-resolved particle growth in nanodusty plasmas},}\ }\href@noop {} {\bibfield  {journal} {\bibinfo  {journal} {J. Phys. D: Appl. Phys.}\ }\textbf {\bibinfo {volume} {48}},\ \bibinfo {pages} {465203} (\bibinfo {year} {2015})}\BibitemShut {NoStop}%
\bibitem [{\citenamefont {Zhang}\ \emph {et~al.}(2014)\citenamefont {Zhang}, \citenamefont {Zhou}, \citenamefont {Liu},\ and\ \citenamefont {Yu}}]{zhang2014new}%
  \BibitemOpen
  \bibfield  {author} {\bibinfo {author} {\bibfnamefont {J.}~\bibnamefont {Zhang}}, \bibinfo {author} {\bibfnamefont {P.}~\bibnamefont {Zhou}}, \bibinfo {author} {\bibfnamefont {J.}~\bibnamefont {Liu}}, \ and\ \bibinfo {author} {\bibfnamefont {J.}~\bibnamefont {Yu}},\ }\bibfield  {title} {\enquote {\bibinfo {title} {New understanding of the difference of photocatalytic activity among anatase, rutile and brookite tio2},}\ }\href@noop {} {\bibfield  {journal} {\bibinfo  {journal} {Phys. Chem. Chem. Phys.}\ }\textbf {\bibinfo {volume} {16}},\ \bibinfo {pages} {20382--20386} (\bibinfo {year} {2014})}\BibitemShut {NoStop}%
\bibitem [{\citenamefont {Luttrell}\ \emph {et~al.}(2014)\citenamefont {Luttrell}, \citenamefont {Halpegamage}, \citenamefont {Tao}, \citenamefont {Kramer}, \citenamefont {Sutter},\ and\ \citenamefont {Batzill}}]{luttrell2014anatase}%
  \BibitemOpen
  \bibfield  {author} {\bibinfo {author} {\bibfnamefont {T.}~\bibnamefont {Luttrell}}, \bibinfo {author} {\bibfnamefont {S.}~\bibnamefont {Halpegamage}}, \bibinfo {author} {\bibfnamefont {J.}~\bibnamefont {Tao}}, \bibinfo {author} {\bibfnamefont {A.}~\bibnamefont {Kramer}}, \bibinfo {author} {\bibfnamefont {E.}~\bibnamefont {Sutter}}, \ and\ \bibinfo {author} {\bibfnamefont {M.}~\bibnamefont {Batzill}},\ }\bibfield  {title} {\enquote {\bibinfo {title} {Why is anatase a better photocatalyst than rutile?-{Model} studies on epitaxial {TiO}$_2$ films},}\ }\href@noop {} {\bibfield  {journal} {\bibinfo  {journal} {Sci. Rep.}\ }\textbf {\bibinfo {volume} {4}},\ \bibinfo {pages} {4043} (\bibinfo {year} {2014})}\BibitemShut {NoStop}%
\bibitem [{\citenamefont {Merlino}(2006)}]{merlino06}%
  \BibitemOpen
  \bibfield  {author} {\bibinfo {author} {\bibfnamefont {R.}~\bibnamefont {Merlino}},\ }\href@noop {} {\emph {\bibinfo {title} {Dusty plasmas and applications in space and industry}}},\ Vol.~\bibinfo {volume} {81}\ (\bibinfo  {publisher} {Transworld Research Network Kerala, India},\ \bibinfo {year} {2006})\ pp.\ \bibinfo {pages} {73--110}\BibitemShut {NoStop}%
\bibitem [{\citenamefont {de~Wetering}\ \emph {et~al.}(2015)\citenamefont {de~Wetering}, \citenamefont {Brooimans}, \citenamefont {Nijdam}, \citenamefont {Beckers},\ and\ \citenamefont {Kroesen}}]{van2015}%
  \BibitemOpen
  \bibfield  {author} {\bibinfo {author} {\bibfnamefont {F.~V.}\ \bibnamefont {de~Wetering}}, \bibinfo {author} {\bibfnamefont {R.}~\bibnamefont {Brooimans}}, \bibinfo {author} {\bibfnamefont {S.}~\bibnamefont {Nijdam}}, \bibinfo {author} {\bibfnamefont {J.}~\bibnamefont {Beckers}}, \ and\ \bibinfo {author} {\bibfnamefont {G.}~\bibnamefont {Kroesen}},\ }\bibfield  {title} {\enquote {\bibinfo {title} {Fast and interrupted expansion in cyclic void growth in dusty plasma},}\ }\href@noop {} {\bibfield  {journal} {\bibinfo  {journal} {J. Phys. D: Appl. Phys.}\ }\textbf {\bibinfo {volume} {48}},\ \bibinfo {pages} {035204} (\bibinfo {year} {2015})}\BibitemShut {NoStop}%
\bibitem [{\citenamefont {Norl{\'e}n}(1973)}]{norlen1973}%
  \BibitemOpen
  \bibfield  {author} {\bibinfo {author} {\bibfnamefont {G.}~\bibnamefont {Norl{\'e}n}},\ }\bibfield  {title} {\enquote {\bibinfo {title} {Wavelengths and energy levels of {Ar I} and {Ar II} based on new interferometric measurements in the {Region} 3 400-9 800 {\aa}},}\ }\href@noop {} {\bibfield  {journal} {\bibinfo  {journal} {Phys. Scr.}\ }\textbf {\bibinfo {volume} {8}},\ \bibinfo {pages} {249} (\bibinfo {year} {1973})}\BibitemShut {NoStop}%
\bibitem [{\citenamefont {Wiese}\ \emph {et~al.}(1989)\citenamefont {Wiese}, \citenamefont {Brault}, \citenamefont {Danzmann}, \citenamefont {Helbig},\ and\ \citenamefont {Kock}}]{wiese1989}%
  \BibitemOpen
  \bibfield  {author} {\bibinfo {author} {\bibfnamefont {W.}~\bibnamefont {Wiese}}, \bibinfo {author} {\bibfnamefont {J.}~\bibnamefont {Brault}}, \bibinfo {author} {\bibfnamefont {K.}~\bibnamefont {Danzmann}}, \bibinfo {author} {\bibfnamefont {V.}~\bibnamefont {Helbig}}, \ and\ \bibinfo {author} {\bibfnamefont {M.}~\bibnamefont {Kock}},\ }\bibfield  {title} {\enquote {\bibinfo {title} {Unified set of atomic transition probabilities for neutral argon},}\ }\href@noop {} {\bibfield  {journal} {\bibinfo  {journal} {Phys. Rev. A}\ }\textbf {\bibinfo {volume} {39}},\ \bibinfo {pages} {2461} (\bibinfo {year} {1989})}\BibitemShut {NoStop}%
\bibitem [{\citenamefont {Kramida}\ \emph {et~al.}(2022)\citenamefont {Kramida}, \citenamefont {{Yu.~Ralchenko}}, \citenamefont {Reader},\ and\ \citenamefont {{and NIST ASD Team}}}]{NIST_ASD}%
  \BibitemOpen
  \bibfield  {author} {\bibinfo {author} {\bibfnamefont {A.}~\bibnamefont {Kramida}}, \bibinfo {author} {\bibnamefont {{Yu.~Ralchenko}}}, \bibinfo {author} {\bibfnamefont {J.}~\bibnamefont {Reader}}, \ and\ \bibinfo {author} {\bibnamefont {{and NIST ASD Team}}},\ }\href@noop {} {}\bibinfo {howpublished} {{NIST Atomic Spectra Database (ver. 5.10), [Online]. Available: {\tt{https://physics.nist.gov/asd}} [2023, September 12]. National Institute of Standards and Technology, Gaithersburg, MD.}} (\bibinfo {year} {2022})\BibitemShut {NoStop}%
\bibitem [{\citenamefont {Balachandran}\ and\ \citenamefont {Eror}(1982)}]{balachandran1982}%
  \BibitemOpen
  \bibfield  {author} {\bibinfo {author} {\bibfnamefont {U.}~\bibnamefont {Balachandran}}\ and\ \bibinfo {author} {\bibfnamefont {N.}~\bibnamefont {Eror}},\ }\bibfield  {title} {\enquote {\bibinfo {title} {Raman spectra of titanium dioxide},}\ }\href@noop {} {\bibfield  {journal} {\bibinfo  {journal} {J. Solid State Chem.}\ }\textbf {\bibinfo {volume} {42}},\ \bibinfo {pages} {276--282} (\bibinfo {year} {1982})}\BibitemShut {NoStop}%
\bibitem [{\citenamefont {Fictorie}, \citenamefont {Evans},\ and\ \citenamefont {Gladfelter}(1994)}]{fictorie1994kinetic}%
  \BibitemOpen
  \bibfield  {author} {\bibinfo {author} {\bibfnamefont {C.~P.}\ \bibnamefont {Fictorie}}, \bibinfo {author} {\bibfnamefont {J.~F.}\ \bibnamefont {Evans}}, \ and\ \bibinfo {author} {\bibfnamefont {W.~L.}\ \bibnamefont {Gladfelter}},\ }\bibfield  {title} {\enquote {\bibinfo {title} {Kinetic and mechanistic study of the chemical vapor deposition of titanium dioxide thin films using tetrakis-(isopropoxo)-titanium {(IV)}},}\ }\href@noop {} {\bibfield  {journal} {\bibinfo  {journal} {J. Vac. Sci. Technol. A}\ }\textbf {\bibinfo {volume} {12}},\ \bibinfo {pages} {1108--1113} (\bibinfo {year} {1994})}\BibitemShut {NoStop}%
\bibitem [{\citenamefont {Ahn}, \citenamefont {Park},\ and\ \citenamefont {Park}(2003)}]{ahn2003kinetic}%
  \BibitemOpen
  \bibfield  {author} {\bibinfo {author} {\bibfnamefont {K.-H.}\ \bibnamefont {Ahn}}, \bibinfo {author} {\bibfnamefont {Y.-B.}\ \bibnamefont {Park}}, \ and\ \bibinfo {author} {\bibfnamefont {D.-W.}\ \bibnamefont {Park}},\ }\bibfield  {title} {\enquote {\bibinfo {title} {Kinetic and mechanistic study on the chemical vapor deposition of titanium dioxide thin films by in situ {FT-IR} using {TTIP}},}\ }\href@noop {} {\bibfield  {journal} {\bibinfo  {journal} {Surf. Coat. Technol.}\ }\textbf {\bibinfo {volume} {171}},\ \bibinfo {pages} {198--204} (\bibinfo {year} {2003})}\BibitemShut {NoStop}%
\bibitem [{\citenamefont {Kroesen}\ \emph {et~al.}(1996)\citenamefont {Kroesen}, \citenamefont {Den~Boer}, \citenamefont {Boufendi}, \citenamefont {Vivet}, \citenamefont {Khouli}, \citenamefont {Bouchoule},\ and\ \citenamefont {De~Hoog}}]{kroesen1996situ}%
  \BibitemOpen
  \bibfield  {author} {\bibinfo {author} {\bibfnamefont {G.}~\bibnamefont {Kroesen}}, \bibinfo {author} {\bibfnamefont {J.}~\bibnamefont {Den~Boer}}, \bibinfo {author} {\bibfnamefont {L.}~\bibnamefont {Boufendi}}, \bibinfo {author} {\bibfnamefont {F.}~\bibnamefont {Vivet}}, \bibinfo {author} {\bibfnamefont {M.}~\bibnamefont {Khouli}}, \bibinfo {author} {\bibfnamefont {A.}~\bibnamefont {Bouchoule}}, \ and\ \bibinfo {author} {\bibfnamefont {F.}~\bibnamefont {De~Hoog}},\ }\bibfield  {title} {\enquote {\bibinfo {title} {In situ infrared absorption spectroscopy of dusty plasmas},}\ }\href@noop {} {\bibfield  {journal} {\bibinfo  {journal} {J. Vac. Sci. Technol. A}\ }\textbf {\bibinfo {volume} {14}},\ \bibinfo {pages} {546--549} (\bibinfo {year} {1996})}\BibitemShut {NoStop}%
\bibitem [{\citenamefont {Deschenaux}\ \emph {et~al.}(1999)\citenamefont {Deschenaux}, \citenamefont {Affolter}, \citenamefont {Magni}, \citenamefont {Hollenstein},\ and\ \citenamefont {Fayet}}]{deschenaux1999investigations}%
  \BibitemOpen
  \bibfield  {author} {\bibinfo {author} {\bibfnamefont {C.}~\bibnamefont {Deschenaux}}, \bibinfo {author} {\bibfnamefont {A.}~\bibnamefont {Affolter}}, \bibinfo {author} {\bibfnamefont {D.}~\bibnamefont {Magni}}, \bibinfo {author} {\bibfnamefont {C.}~\bibnamefont {Hollenstein}}, \ and\ \bibinfo {author} {\bibfnamefont {P.}~\bibnamefont {Fayet}},\ }\bibfield  {title} {\enquote {\bibinfo {title} {Investigations of {CH}$_4$, {C$_2$H$_2$} and {C$_2$H$_4$} dusty rf plasmas by means of {FTIR} absorption spectroscopy and mass spectrometry},}\ }\href@noop {} {\bibfield  {journal} {\bibinfo  {journal} {J. Phys. D: Appl. Phys.}\ }\textbf {\bibinfo {volume} {32}},\ \bibinfo {pages} {1876} (\bibinfo {year} {1999})}\BibitemShut {NoStop}%
\bibitem [{\citenamefont {Ouaras}\ \emph {et~al.}(2014)\citenamefont {Ouaras}, \citenamefont {Delacqua}, \citenamefont {Lombardi}, \citenamefont {R{\"o}pcke}, \citenamefont {Wartel}, \citenamefont {Bonnin}, \citenamefont {Redolfi},\ and\ \citenamefont {Hassouni}}]{ouaras2014situ}%
  \BibitemOpen
  \bibfield  {author} {\bibinfo {author} {\bibfnamefont {K.}~\bibnamefont {Ouaras}}, \bibinfo {author} {\bibfnamefont {L.~C.}\ \bibnamefont {Delacqua}}, \bibinfo {author} {\bibfnamefont {G.}~\bibnamefont {Lombardi}}, \bibinfo {author} {\bibfnamefont {J.}~\bibnamefont {R{\"o}pcke}}, \bibinfo {author} {\bibfnamefont {M.}~\bibnamefont {Wartel}}, \bibinfo {author} {\bibfnamefont {X.}~\bibnamefont {Bonnin}}, \bibinfo {author} {\bibfnamefont {M.}~\bibnamefont {Redolfi}}, \ and\ \bibinfo {author} {\bibfnamefont {K.}~\bibnamefont {Hassouni}},\ }\bibfield  {title} {\enquote {\bibinfo {title} {In-situ diagnostics of hydrocarbon dusty plasmas using quantum cascade laser absorption spectroscopy and mass spectrometry},}\ }\href@noop {} {\bibfield  {journal} {\bibinfo  {journal} {J. Plasma Phys.}\ }\textbf {\bibinfo {volume} {80}},\ \bibinfo {pages} {833--841} (\bibinfo {year} {2014})}\BibitemShut {NoStop}%
\bibitem [{\citenamefont {Denysenko}\ \emph {et~al.}(2018)\citenamefont {Denysenko}, \citenamefont {von Wahl}, \citenamefont {Labidi}, \citenamefont {Mikikian}, \citenamefont {Kersten}, \citenamefont {Gibert}, \citenamefont {Kovacevic},\ and\ \citenamefont {Azarenkov}}]{denysenko2018modeling}%
  \BibitemOpen
  \bibfield  {author} {\bibinfo {author} {\bibfnamefont {I.}~\bibnamefont {Denysenko}}, \bibinfo {author} {\bibfnamefont {E.}~\bibnamefont {von Wahl}}, \bibinfo {author} {\bibfnamefont {S.}~\bibnamefont {Labidi}}, \bibinfo {author} {\bibfnamefont {M.}~\bibnamefont {Mikikian}}, \bibinfo {author} {\bibfnamefont {H.}~\bibnamefont {Kersten}}, \bibinfo {author} {\bibfnamefont {T.}~\bibnamefont {Gibert}}, \bibinfo {author} {\bibfnamefont {E.}~\bibnamefont {Kovacevic}}, \ and\ \bibinfo {author} {\bibfnamefont {N.~A.}\ \bibnamefont {Azarenkov}},\ }\bibfield  {title} {\enquote {\bibinfo {title} {Modeling of argon--acetylene dusty plasma},}\ }\href@noop {} {\bibfield  {journal} {\bibinfo  {journal} {Plasma Phys. Controlled Fusion}\ }\textbf {\bibinfo {volume} {61}},\ \bibinfo {pages} {014014} (\bibinfo {year} {2018})}\BibitemShut {NoStop}%
\bibitem [{\citenamefont {Stefanovic}\ \emph {et~al.}(2003)\citenamefont {Stefanovic}, \citenamefont {Kovacevic}, \citenamefont {J.Berndt},\ and\ \citenamefont {Winter}}]{stefanovic03}%
  \BibitemOpen
  \bibfield  {author} {\bibinfo {author} {\bibfnamefont {I.}~\bibnamefont {Stefanovic}}, \bibinfo {author} {\bibfnamefont {E.}~\bibnamefont {Kovacevic}}, \bibinfo {author} {\bibnamefont {J.Berndt}}, \ and\ \bibinfo {author} {\bibfnamefont {J.}~\bibnamefont {Winter}},\ }\bibfield  {title} {\enquote {\bibinfo {title} {H$\alpha$ emission in the presence of dust in an {Ar}-{C$_2$H$_2$} radio-frequency discharge},}\ }\href@noop {} {\bibfield  {journal} {\bibinfo  {journal} {New J. Phys.}\ }\textbf {\bibinfo {volume} {5}},\ \bibinfo {pages} {39} (\bibinfo {year} {2003})}\BibitemShut {NoStop}%
\bibitem [{\citenamefont {Donders}, \citenamefont {Staps},\ and\ \citenamefont {Beckers}(2022)}]{Donders22}%
  \BibitemOpen
  \bibfield  {author} {\bibinfo {author} {\bibfnamefont {T.}~\bibnamefont {Donders}}, \bibinfo {author} {\bibfnamefont {T.}~\bibnamefont {Staps}}, \ and\ \bibinfo {author} {\bibfnamefont {J.}~\bibnamefont {Beckers}},\ }\bibfield  {title} {\enquote {\bibinfo {title} {Characterization of cyclic dust growth in a low-pressure, radio-frequency driven argon-hexamethyldisiloxane plasma},}\ }\href@noop {} {\bibfield  {journal} {\bibinfo  {journal} {J. Phys. D: Appl. Phys.}\ }\textbf {\bibinfo {volume} {55}},\ \bibinfo {pages} {1--15} (\bibinfo {year} {2022})}\BibitemShut {NoStop}%
\bibitem [{\citenamefont {Fridman}\ \emph {et~al.}(1996)\citenamefont {Fridman}, \citenamefont {Boufendi}, \citenamefont {Hbid}, \citenamefont {Potapkin},\ and\ \citenamefont {Bouchoule}}]{Fridman96}%
  \BibitemOpen
  \bibfield  {author} {\bibinfo {author} {\bibfnamefont {A.}~\bibnamefont {Fridman}}, \bibinfo {author} {\bibfnamefont {L.}~\bibnamefont {Boufendi}}, \bibinfo {author} {\bibfnamefont {T.}~\bibnamefont {Hbid}}, \bibinfo {author} {\bibfnamefont {B.}~\bibnamefont {Potapkin}}, \ and\ \bibinfo {author} {\bibfnamefont {A.}~\bibnamefont {Bouchoule}},\ }\bibfield  {title} {\enquote {\bibinfo {title} {Dusty plasma formation: {Physics} and critical phenomena. {Theoretical} approach},}\ }\href@noop {} {\bibfield  {journal} {\bibinfo  {journal} {J. Appl. Phys.}\ }\textbf {\bibinfo {volume} {79}},\ \bibinfo {pages} {1303--1314} (\bibinfo {year} {1996})}\BibitemShut {NoStop}%
\bibitem [{\citenamefont {Shi}, \citenamefont {Elvati},\ and\ \citenamefont {Violi}(2021)}]{shi2021growth}%
  \BibitemOpen
  \bibfield  {author} {\bibinfo {author} {\bibfnamefont {X.}~\bibnamefont {Shi}}, \bibinfo {author} {\bibfnamefont {P.}~\bibnamefont {Elvati}}, \ and\ \bibinfo {author} {\bibfnamefont {A.}~\bibnamefont {Violi}},\ }\bibfield  {title} {\enquote {\bibinfo {title} {On the growth of {Si} nanoparticles in non-thermal plasma: physisorption to chemisorption conversion},}\ }\href@noop {} {\bibfield  {journal} {\bibinfo  {journal} {J. Phys. D: Appl. Phys.}\ }\textbf {\bibinfo {volume} {54}},\ \bibinfo {pages} {365203} (\bibinfo {year} {2021})}\BibitemShut {NoStop}%
\bibitem [{\citenamefont {Winter}(2004)}]{winter2004dust}%
  \BibitemOpen
  \bibfield  {author} {\bibinfo {author} {\bibfnamefont {J.}~\bibnamefont {Winter}},\ }\bibfield  {title} {\enquote {\bibinfo {title} {Dust in fusion devices—a multi-faceted problem connecting high-and low-temperature plasma physics},}\ }\href@noop {} {\bibfield  {journal} {\bibinfo  {journal} {Plasma Phys. Controlled Fusion}\ }\textbf {\bibinfo {volume} {46}},\ \bibinfo {pages} {B583} (\bibinfo {year} {2004})}\BibitemShut {NoStop}%
\end{thebibliography}%

\end{document}